\documentstyle[a4p,12pt,epsfig,cite,rotating]{article}

\begin{document}
%========================================================================
%
%  define a few new names (remember math mode ne text mode!)
%
\newcommand{\xE}{\mbox{$x_E$}}
\newcommand{\xp}{\mbox{$x_p$}}
\newcommand{\lnxp}{\mbox{$\log(1/x_p)$}}
\newcommand{\costh}{\mbox{$\cos\theta$}}
\newcommand{\degree}{$^{\circ}\space$}
\newcommand{\cth}    {$|\cos\theta|$}
\newcommand{\Esh}    {$E_{\mbox{\scriptsize{shower}}}$}
\newcommand{\Res}    {$\Sigma_{|\mbox{\scriptsize{Residuals}}|}/E_d$}
%
%--   ... Pi0 prob
%
\newcommand{\Pg} {\mbox{$P_{\gamma}$}}
\newcommand{\Ppi} {\mbox{$P_{\gamma\gamma}$}}
\newcommand{\Ppim} {\mbox{$G_{\gamma\gamma}(m)$}}
\newcommand{\Ptpi} {\mbox{$\tilde{P}_{\gamma\gamma}$}}
\newcommand{\Ptpim} {\mbox{$\tilde{G}_{\gamma\gamma}(m)$}}
%                                            Pi0 prob, with indexes
\newcommand{\iPpi} {\mbox{$P_{\gamma_i\gamma_j}$}}
\newcommand{\iPpim} {\mbox{$G_{\gamma_i\gamma_j}(m)$}}
\newcommand{\iPtpi} {\mbox{$\tilde{P}_{\gamma_i\gamma_j}$}}
\newcommand{\iPtpim} {\mbox{$\tilde{G}_{\gamma_i\gamma_j}^{(m)}$}}
%                                            P slash always with indexes
\newcommand{\Pslash} {\mbox{$P\!\!\!\!/_{\gamma_i\gamma_j}$}}
%
%  ... The results
%
\newcommand{\gamresl} {\mbox{20.97 $\pm$ 0.02 $\pm$ 1.07 $\pm$ 0.42 }}
\newcommand{\gamres}  {\mbox{20.97 $\pm$ 0.02 $\pm$ 1.15 }}
\newcommand{\gamrs}   {\mbox{20.97 $\pm$ 1.15 }}
\newcommand{\pizresl}  {\mbox{9.55 $\pm$ 0.06 $\pm$ 0.72 $\pm$ 0.21 }}
\newcommand{\pizres}   {\mbox{9.55 $\pm$ 0.06 $\pm$ 0.75 }}
\newcommand{\pizrs}    {\mbox{9.55 $\pm$ 0.76 }}
\newcommand{\etaresl}  {\mbox{0.97 $\pm$ 0.03 $\pm$ 0.10 $\pm$ 0.04 }}
\newcommand{\etares}   {\mbox{0.97 $\pm$ 0.03 $\pm$ 0.11 }}
\newcommand{\etars}    {\mbox{0.97 $\pm$ 0.11 }}
\newcommand{\rhoresl}  {\mbox{2.40 $\pm$ 0.06 $\pm$ 0.43 $\pm$ 0.02 }}
\newcommand{\rhores}   {\mbox{2.40 $\pm$ 0.06 $\pm$ 0.43 }}
\newcommand{\rhors}    {\mbox{2.40 $\pm$ 0.44 }}
\newcommand{\omeresl}  {\mbox{1.04 $\pm$ 0.04 $\pm$ 0.13 $\pm$ 0.03 }}
\newcommand{\omeres}   {\mbox{1.04 $\pm$ 0.04 $\pm$ 0.14 }}
\newcommand{\omers}    {\mbox{1.04 $\pm$ 0.14 }}
\newcommand{\etpresl}  {\mbox{0.14 $\pm$ 0.01 $\pm$ 0.02 $\pm$ 0.01 }}
\newcommand{\etpres}   {\mbox{0.14 $\pm$ 0.01 $\pm$ 0.02 }}
\newcommand{\etprs}    {\mbox{0.14 $\pm$ 0.03 }}
\newcommand{\azeresl}  {\mbox{0.27 $\pm$ 0.04 $\pm$ 0.10 $\pm$ 0.01 }}
\newcommand{\azeres}   {\mbox{0.27 $\pm$ 0.04 $\pm$ 0.10 }}
\newcommand{\azers}    {\mbox{0.27 $\pm$ 0.11 }}
%
%--    ... Results for x_p > 0.1
%
\newcommand{\etaresh}  {\mbox{0.344 $\pm$ 0.030 }}
\newcommand{\etpresh}  {\mbox{0.069 $\pm$ 0.012 }}
%
%--    ... Useful numbers
%
\newcommand{\gamerrpct}      {4.7}
\newcommand{\gamcerrpct}     {8.3}
\newcommand{\piggerrpct}     {8.4}
\newcommand{\piggcerrpct}   {12.2}
\newcommand{\pigcgcerrpct}  {23.6}
\newcommand{\etaggerrpct}   {14.0}
\newcommand{\etaggcerrpct}  {34.3}
\newcommand{\etaggterrpct}  {13.9}
\newcommand{\etappperrpct}  {12.9}
%
%--    ... Particles
%
\newcommand{\epem}{\mbox{$\mathrm{e^+e^-}$}}
\newcommand{\Zzero}{\mbox{${\mathrm{Z}^0}$}}
\newcommand{\azpm}{\mbox{$\mathrm{a}_0^{\pm}$}}
\newcommand{\rpm}{\mbox{$\rho^{\pm}$}}
\newcommand{\piz}{\mbox{${\pi^0}$}}
\newcommand{\pipm}{\mbox{$\pi^{\pm}$}}
%
%--   ... Decay channels
%
\newcommand {\threepi} {\mbox{$\pi^0\pi^+\pi^-$}}
\newcommand {\pigg} {$\pi^0 \rightarrow \gamma\gamma$}
\newcommand {\piggc} {$\pi^0 \rightarrow \gamma\gamma_c$}
\newcommand {\pigcgc} {$\pi^0 \rightarrow \gamma_c\gamma_c$}
\newcommand {\etagg} {$\eta \rightarrow \gamma\gamma$}
\newcommand {\etaggc} {$\eta \rightarrow \gamma\gamma_c$}
\newcommand {\etagcgc} {$\eta \rightarrow \gamma_c\gamma_c$}
\newcommand {\etapgg} {$\eta ' \rightarrow \gamma\gamma$}
\newcommand {\rhopipi} {$\rho^{\pm} \rightarrow \pi^0\pi^{\pm}$}
\newcommand {\omegapipipi} {$\omega \rightarrow \pi^0\pi^+\pi^-$}
\newcommand {\etapipipi} {$\eta \rightarrow \pi^0\pi^+\pi^-$}
\newcommand {\azetapi} {$\mathrm{a}_0^{\pm} \rightarrow \eta\pi^{\pm}$}
\newcommand {\etapetapipi} {$\eta ' \rightarrow \eta\pi^+\pi^-$}
%
%----------------------
 
%----- Title page -----
%----------------------
\begin{titlepage}

\begin{center}
{\large   EUROPEAN LABORATORY FOR PARTICLE PHYSICS }
\end{center}
\bigskip
\begin{flushright}
       CERN-EP/98-054 \\
%       \today \\
       April 14, 1998
\end{flushright}

\vfill

\begin{center}
{\LARGE\bf  Photon and Light Meson Production
            in Hadronic Z$^0$ Decays }
\end{center}

\bigskip\bigskip\bigskip
\begin{center}
{\LARGE The OPAL Collaboration}
\end{center}

\vfill

\begin{center}{\large\bf Abstract}
\end{center}

The inclusive production rates and differential cross-sections of
photons and mesons with a final state containing photons
have been measured with the OPAL detector at LEP.
The light mesons covered by the measurements are
the \piz, $\eta$, $\rho(770)^{\pm}$, $\omega(782)$,
$\eta'(958)$ and a$_0(980)^{\pm}$.
The particle multiplicities per hadronic \Zzero\  decay,
extrapolated to the full energy range, are:
\begin{eqnarray}
\langle n_{\gamma} \rangle              & = & \gamres, \nonumber \\
\langle n_{\pi^0} \rangle               & = & \pizres, \nonumber \\
\langle n_{\eta} \rangle                & = & \etares, \nonumber \\
\langle n_{\rho^{\pm}} \rangle          & = & \rhores, \nonumber \\
\langle n_{\omega} \rangle              & = & \omeres, \nonumber \\
\langle n_{\eta'} \rangle               & = & \etpres, \nonumber \\
\langle n_{{\mathrm a}_0^{\pm}} \rangle & = & \azeres, \nonumber
\end{eqnarray}
where the first errors are statistical and the second systematic.
In general, the results are in agreement with the
predictions of the JETSET and HERWIG Monte Carlo models.

\vfill

\begin{center}
 {\large (Submitted to European Physical Journal, C)}
\end{center}
 
\end{titlepage}
%%%%%%%%%%%%%%%%%%%%%%%%%%%%%%%%%
\begin{center}{\Large        The OPAL Collaboration
}\end{center}\bigskip
\begin{center}{
%begin authorlist
K.\thinspace Ackerstaff$^{  8}$,
G.\thinspace Alexander$^{ 23}$,
J.\thinspace Allison$^{ 16}$,
N.\thinspace Altekamp$^{  5}$,
K.J.\thinspace Anderson$^{  9}$,
S.\thinspace Anderson$^{ 12}$,
S.\thinspace Arcelli$^{  2}$,
S.\thinspace Asai$^{ 24}$,
S.F.\thinspace Ashby$^{  1}$,
D.\thinspace Axen$^{ 29}$
% ...
%end authorlist
}\end{center}\bigskip
\bigskip
%begin institutes
$^{  1}$School of Physics and Astronomy, University of Birmingham,
Birmingham B15 2TT, UK
\newline
$^{  2}$Dipartimento di Fisica dell' Universit\`a di Bologna and INFN,
I-40126 Bologna, Italy
\newline
$^{  3}$Physikalisches Institut, Universit\"at Bonn,
D-53115 Bonn, Germany
\newline
$^{  4}$Department of Physics, University of California,
Riverside CA 92521, USA
\newline
$^{  5}$Cavendish Laboratory, Cambridge CB3 0HE, UK
\newline
$^{  6}$Ottawa-Carleton Institute for Physics,
Department of Physics, Carleton University,
Ottawa, Ontario K1S 5B6, Canada
\newline
$^{  7}$Centre for Research in Particle Physics,
Carleton University, Ottawa, Ontario K1S 5B6, Canada
\newline
$^{  8}$CERN, European Organisation for Particle Physics,
CH-1211 Geneva 23, Switzerland
\newline
$^{  9}$Enrico Fermi Institute and Department of Physics,
University of Chicago, Chicago IL 60637, USA
\newline
$^{ 10}$Fakult\"at f\"ur Physik, Albert Ludwigs Universit\"at,
D-79104 Freiburg, Germany
\newline
$^{ 11}$Physikalisches Institut, Universit\"at
Heidelberg, D-69120 Heidelberg, Germany
\newline
$^{ 12}$Indiana University, Department of Physics,
Swain Hall West 117, Bloomington IN 47405, USA
\newline
$^{ 13}$Queen Mary and Westfield College, University of London,
London E1 4NS, UK
\newline
$^{ 14}$Technische Hochschule Aachen, III Physikalisches Institut,
Sommerfeldstrasse 26-28, D-52056 Aachen, Germany
\newline
$^{ 15}$University College London, London WC1E 6BT, UK
\newline
$^{ 16}$Department of Physics, Schuster Laboratory, The University,
Manchester M13 9PL, UK
\newline
$^{ 17}$Department of Physics, University of Maryland,
College Park, MD 20742, USA
\newline
$^{ 18}$Laboratoire de Physique Nucl\'eaire, Universit\'e de Montr\'eal,
Montr\'eal, Quebec H3C 3J7, Canada
\newline
$^{ 19}$University of Oregon, Department of Physics, Eugene
OR 97403, USA
\newline
$^{ 20}$Rutherford Appleton Laboratory, Chilton,
Didcot, Oxfordshire OX11 0QX, UK
\newline
$^{ 22}$Department of Physics, Technion-Israel Institute of
Technology, Haifa 32000, Israel
\newline
$^{ 23}$Department of Physics and Astronomy, Tel Aviv University,
Tel Aviv 69978, Israel
\newline
$^{ 24}$International Centre for Elementary Particle Physics and
Department of Physics, University of Tokyo, Tokyo 113, and
Kobe University, Kobe 657, Japan
\newline
$^{ 25}$Institute of Physical and Environmental Sciences,
Brunel University, Uxbridge, Middlesex UB8 3PH, UK
\newline
$^{ 26}$Particle Physics Department, Weizmann Institute of Science,
Rehovot 76100, Israel
\newline
$^{ 27}$Universit\"at Hamburg/DESY, II Institut f\"ur Experimental
Physik, Notkestrasse 85, D-22607 Hamburg, Germany
\newline
$^{ 28}$University of Victoria, Department of Physics, P O Box 3055,
Victoria BC V8W 3P6, Canada
\newline
$^{ 29}$University of British Columbia, Department of Physics,
Vancouver BC V6T 1Z1, Canada
\newline
$^{ 30}$University of Alberta,  Department of Physics,
Edmonton AB T6G 2J1, Canada
\newline
$^{ 31}$Duke University, Dept of Physics,
Durham, NC 27708-0305, USA
\newline
$^{ 32}$Research Institute for Particle and Nuclear Physics,
H-1525 Budapest, P O  Box 49, Hungary
\newline
$^{ 33}$Institute of Nuclear Research,
H-4001 Debrecen, P O  Box 51, Hungary
\newline
$^{ 34}$Ludwigs-Maximilians-Universit\"at M\"unchen,
Sektion Physik, Am Coulombwall 1, D-85748 Garching, Germany
\newline
%end institutes
\bigskip\newline
%begin notes
$^{  a}$ and at TRIUMF, Vancouver, Canada V6T 2A3
\newline
$^{  b}$ and Royal Society University Research Fellow
\newline
$^{  c}$ and Institute of Nuclear Research, Debrecen, Hungary
\newline
$^{  d}$ and Department of Experimental Physics, Lajos Kossuth
University, Debrecen, Hungary
\newline
$^{  e}$ on leave of absence from the University of Freiburg
\newline
%end notes
%%%%%%%%%%%%%%%%%%%%%%%%%%%%%%%%%

%========================================================================
 
%-------------------------
%----- Document text -----
%-------------------------

\newpage

%%%%%%%%%%%%%%%%%%%%%%%%%%%%%%%%%%%%%%%%%%%%%%%%%%%%%%%%%%%%%%%%%%%%%%%%%%%%

\section{Introduction}
\label{sect-intro}

In high-energy collisions,
the transition from quarks and gluons to stable hadrons
can only be described by phenomenological models~\cite{bib-had}.
Among the basic features that these hadronisation models attempt to
reproduce are the multiplicity and energy spectrum of
each hadron species.
The large sample of hadronic \Zzero\  decays collected at LEP is
ideal to test these models and to improve their accuracy, as the
initial state in this process is theoretically well understood.
With the versatility of the LEP detectors, these measurements
can be extended to most of the low-lying particle
states~\cite{bib-opalpkpi,bib-opalkz,bib-opalphi,bib-opalfz,bib-alcomp,%
bib-delpkpiphi,bib-delrho,bib-delpiz,bib-lpiz,bib-leta,bib-letap}.

This paper describes the OPAL measurements of the differential
production cross-sections in hadronic \Zzero\  decays
of photons and of light mesons
decaying to final states containing at least one photon.
The mesons studied are the \piz, $\eta$,
$\rho(770)^{\pm}$, $\omega(782)$, $\eta'(958)$ and a$_0(980)^{\pm}$.
The measurements require a good understanding of both the
detector response to photons and of the environment in which
these particles are produced. 
For this reason, the results obtained with photons detected as
energy deposits in the electromagnetic calorimeter and as
pairs of tracks from photon conversions to \epem\   in the central
drift chamber are first compared and then combined.
This comprehensive study of the production of mesons
decaying to photons leads to a better
understanding of the systematic effects related to photon
detection, making possible a reliable measurement of the inclusive
production of photons in \Zzero\  decays in a wide energy
range\footnote{For an inclusive measurement of {\em prompt}
photons, excluding hadron decays and initial state radiation,
see ref.~\cite{bib-piotr}.}.
The production cross-section of each particle is presented as a function of its
scaled energy \xE\  = $E_{\mathrm{particle}}/E_{\mathrm{beam}}$
and of \lnxp, where $x_p$ = $p_{\mathrm{particle}}/p_{\mathrm{beam}}$
is the scaled momentum.

Current measurements of photon and \piz, $\eta$,
$\omega$ and $\eta'$ meson production at LEP are limited
by experimental systematic
errors~\cite{bib-alcomp,bib-delpiz,bib-lpiz,bib-leta,bib-letap}.
Compared to these studies, the present measurements cover
a larger fraction of the total rate of these particles,
and a number of the sources of systematic error are different.
This is the first measurement of the inclusive production of
\rpm\  and a$_0^{\pm}$ in high-energy \epem\   collisions.

The outline of the paper is the following. The OPAL detector is
briefly presented, followed by the description
of the event selection and simulation.
The following three sections describe the three steps in
the particle reconstruction.
First, photons are detected either as localised
energy deposits in the electromagnetic calorimeter or as two tracks
from a $\gamma\rightarrow$ \epem\  conversion within the volume of the
central drift chambers.
In the second step, photons are combined in pairs to form \piz\
and $\eta$ meson candidates.
In a final step, the \piz\  and $\eta$ meson candidates
are combined with one charged track or two oppositely
charged tracks to reconstruct \etapipipi, \rhopipi,
\omegapipipi, \etapetapipi\   and \azetapi\  decays.
Each step is described, together with the corresponding rate
measurements and evaluations of systematic errors.
The following section describes the combination of the results
for those particles where more than one decay mode is used.
The resulting differential cross-sections for photons and light
mesons are then compared to the predictions of different
models and to other measurements at LEP.
This is followed by the conclusion.

%%%%%%%%%%%%%%%%%%%%%%%%%%%%%%%%%%%%%%%%%%%%%%%%%%%%%%%%%%%%%%%%%%%%%%%%%%%%

\section{The OPAL detector}
\label{sect-detect}

The OPAL detector and its performance are described in detail
elsewhere~\cite{bib-opaldet}.
Only detector elements of importance for this analysis are
described here.
The central tracking system consists of
three drift chambers which surround a silicon microvertex
detector~\cite{bib-opalsi}, all
within an axial magnetic field of 0.435~T.
The silicon microvertex detector
has two layers, at radii of 6.1 and 7.5~cm from the beam axis,
with an intrinsic resolution of 5~$\mu$m in the
$r-\phi$ plane\footnote{
The OPAL coordinate system is defined so that
$z$~is the coordinate parallel to the e$^-$ beam,
$r$~is the coordinate normal to this axis,
$\theta$~is the polar angle with respect to $z$ and
$\phi$~is the azimuthal angle about the $z$-axis.}.
A precision vertex drift chamber with 24~cm outer radius
provides space points with a resolution of
about 50~$\mu$m in $r-\phi$ and 1 mm in $z$.
The jet chamber has an outer radius 185~cm and
provides up to 159 measurements of space points per track,
with a resolution in the $r-\phi$ plane of about 130~$\mu$m.
The resolution of the $r-\phi$ component of the track momentum
($p_t$) is $\sigma_{p_t}/p_t$ $\sim$ $\sqrt{(0.02)^2+(0.0015p_t)^2}$,
where $p_t$ is in GeV/$c$.
In addition, charged particles can be identified by their
specific ionisation energy loss (d$E$/d$x$)~\cite{bib-dedx}.
The jet chamber is surrounded by a system of $z$-chambers,
thin drift chambers with a resolution of about 300~$\mu$m in
the $z$ coordinate,
which serves to improve the determination of $\theta$.

The tracking detectors and the magnet coil are surrounded by
electromagnetic and hadronic calorimeters
and muon chambers.
In this work, the identification of photons is performed within
the acceptance of the barrel electromagnetic calorimeter.
This consists of a cylindrical array of 9440 lead glass blocks
of 24.6~radiation lengths thickness that
covers the polar angle range $|\cos\theta|<0.82$.
Each block subtends approximately 40$\times$40 mrad$^2$.
The energy resolution is improved by correcting for losses due to
showers initiated in the material in front of the calorimeter.
These showers are detected by thin presampler gas detectors
covering the front surface of the electromagnetic calorimeter.
Time-of-flight scintillators, situated between the magnet coil
and the presampler in the polar angle range $|\cos\theta|<0.72$
are also used to detect these showers.

%%%%%%%%%%%%%%%%%%%%%%%%%%%%%%%%%%%%%%%%%%%%%%%%%%%%%%%%%%%%%%%%%%%%%%%%%%%%

\section{Data selection and event simulation}
\label{sect-event}

This study is based on a sample of 4.1 million hadronic \Zzero\
decays collected by the OPAL detector at LEP at centre-of-mass energies
within $\pm$2 GeV of the \Zzero\  peak.
The hadronic event selection~\cite{bib-opalline} has an efficiency of
98.4 $\pm$ 0.4\% with a background of less than 0.2\%.

The detection efficiencies for the particles under study
are evaluated using 6.4 million hadronic \Zzero\   decays
simulated using the Monte Carlo programs JETSET 7.3 and 7.4~\cite{bib-jetset}
tuned to reproduce the global features of hadronic events as
observed at LEP~\cite{bib-tunenew,bib-tuneold}.
Samples generated by the HERWIG 5.9 program~\cite{bib-tunenew,bib-herwig}
are also used for comparison.
The generated events are passed through a full simulation of the OPAL
detector~\cite{bib-gopal} 
and are subjected to the same event reconstruction and selection
as the data.

%%%%%%%%%%%%%%%%%%%%%%%%%%%%%%%%%%%%%%%%%%%%%%%%%%%%%%%%%%%%%%%%%%%%%%%%%%%%
                                                           
\section{ Photons }
\label{sect-photon}

\subsection{Photons reconstructed using the electromagnetic calorimeter}
\label{sect-recgam}

The electromagnetic calorimeter provides the largest part of
the photon sample.
To resolve a maximum number of photons in the dense
environment of hadronic jets,
the location and the energy of the electromagnetic showers
are obtained from a fit to the energy deposited in
the individual lead glass blocks.
The fit uses a description of the lateral shower profile
as the sum of two exponentials (see for example ref.~\cite{bib-maxent})
and allows a proper treatment of overlapping showers.
The fit can also handle the cases where most of the photon energy is
in a single block, a common occurrence for photons in the
energy range from 0.1 to 0.3 GeV.

Not all energy deposits in the electromagnetic calorimeter
are due to photons.
Many are due to ionisation or to
electromagnetic and hadronic
showers caused by charged particles.
For this reason, a block lying close to the extrapolated path
of a charged track is given a small weight in the fit,
provided its energy does not significantly exceed
the expectation for a hadronic shower.

Photons may lose energy (typically about 0.2 GeV) by initiating
an electromagnetic shower before reaching the calorimeter.
The photon energy is therefore corrected using the signal
recorded in the presampler.
The efficiency for detecting these showers is further increased
by also using the presence of signals in the time-of-flight detectors.

A shower is retained as a photon candidate if it has at least 0.1 GeV
energy  in the lead glass calorimeter and if the energy corrected
using the presampler and the time-of-flight detectors is at least
0.15 GeV.
An acceptance cut of $|\cos\theta|<0.75$ is imposed
to improve the homogeneity in the amount of material
traversed by the photons before reaching the calorimeter.
The momentum direction of the photon is evaluated
assuming that it originates from the primary event
vertex determined as described in~\cite{bib-primvtx}.

According to the Monte Carlo simulations, the overall efficiency for
photons above 0.15~GeV is 69\% within the
acceptance $|\cos\theta|<0.75$.
The purity of the sample is 52\%,
with the most important background being due to
energy deposits from charged particles, neutrons and
K$^0_{{\mathrm L}}$ mesons.
In simulations of hadronic \Zzero\  decays, the photon
angular resolution is approximately 10 mrad and
the energy resolution varies from 30\% at 0.15 GeV to
8\% at 20 GeV.

\begin{figure}[tbp]
\epsfig{file=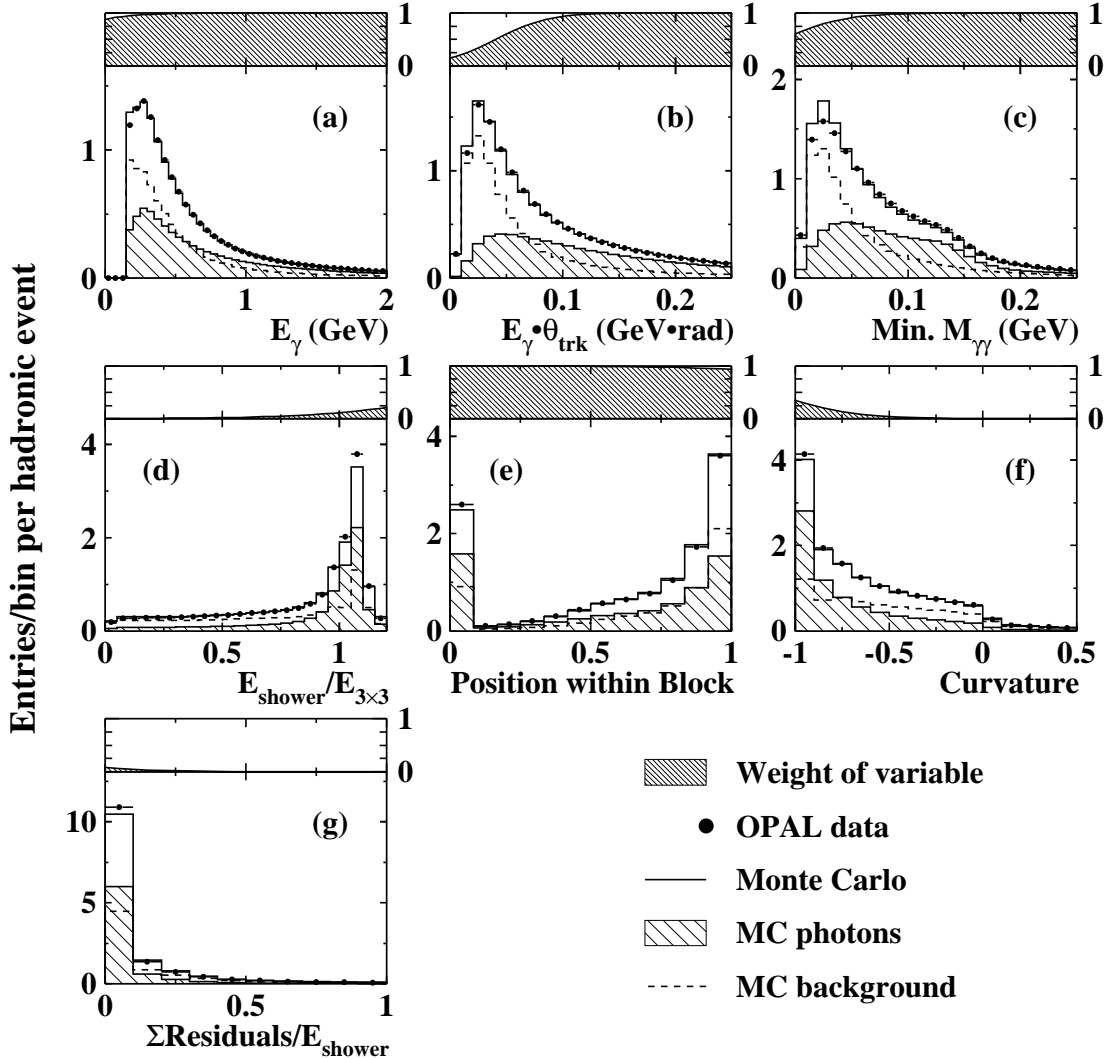
,height=17cm,bbllx=0cm,bblly=0cm,bburx=21cm,bbury=21cm}
\caption  { \label{fig-gvar} 
Distribution of variables used for the discrimination
of photons recorded in the electromagnetic calorimeter.
The points represent the data and the histogram
represents the Monte Carlo simulation, normalised to one event.
The Monte Carlo photons and background are shown
as hatched and dashed histograms, respectively.
The insert shows the weights used in eq.~\protect\ref{eq-acti}.
The variables are:
a) the photon energy;
b) the photon energy multiplied by the angle to the closest charged
   track;
c) the minimum of the invariant mass of the photon with any other
   photon;
d) the fitted shower energy divided by the sum of the energy
   in the 3$\times$3 array of blocks around the shower;
e) the distance of the shower relative to the centre
   of the block, a value of 1 corresponding to the edge
   of the block;
f) the maximum value of $(E_{-1}-E_{0})(E_{+1}-E_{0})/E^2_0$,
   where $E_{-1}$, $E_{0}$, $E_{+1}$ are the energies deposited
   in 3 consecutive blocks in either $\theta$ or $\phi$,
   the index 0 corresponding to the block where the shower
   is centered; and
g) the sum of the residual of the shower fit in the 3$\times$3 array
   of blocks around the shower divided by the shower energy.
}
\end{figure}

For the rate determination, it is useful to compare results
obtained from samples with different purities.
As shown in fig.~\ref{fig-gvar}, the purity depends on the photon energy,
the energy deposited in its vicinity, its distance to the closest charged track,
the shower shape and the quality of the shower fit.
The seven variables shown in fig.~\ref{fig-gvar} 
are combined in one variable $S$:
\begin{eqnarray} \label{eq-acti}
S & = & \frac{1}{N} \sum_{i=1}^{N} ( 1 + \exp((v_i-c_i)/t_i) )^{-1} \mbox{ ,}
\end{eqnarray}
where the index $i$ runs over the $N$=7 variables $v_i$
and the parameters $c_i$ and $t_i$ are chosen such that
the power of the variable $S$
to discriminate between signal and background is maximal.
This is achieved by minimising the ratio
\begin{eqnarray} \label{eq-fish}
 \cal{R}    & = & \frac{\sigma_{s}^2+\sigma_{b}^2}
                         {(\mu_{s}-\mu_{b})^2}  \mbox{ ,}
\end{eqnarray}
where $\mu_{s}$ and $\sigma_{s}$ are the average and rms
values of $S$ for the signal and $\mu_{b}$ and $\sigma_{b}$
are the corresponding values for the background.
The minimisation is performed
using MINUIT~\cite{bib-minuit}
and a sample of simulated events.
The contributions to $S$ of the input variables are shown
as inserts in figs.~\ref{fig-gvar}a to~\ref{fig-gvar}g.

The variable $S$ can be interpreted as the output of
a simplified artificial neural network,
where the number of parameters optimised using a
reference sample of simulated events is
reduced to two per input variable.
In this way, $S$ is forced to depend on the global
properties of the input variables, which should reduce problems due
to the imperfection of the detector simulation.
In the present application, the discrimination losses
relative to more complex artificial networks are negligible.
In contrast to a likelihood method, correlations between the
input variables are partially taken into account by the simultaneous
optimisation of all the $c_i$ and $t_i$ parameters.

\begin{figure}[tbp]
\epsfig{file=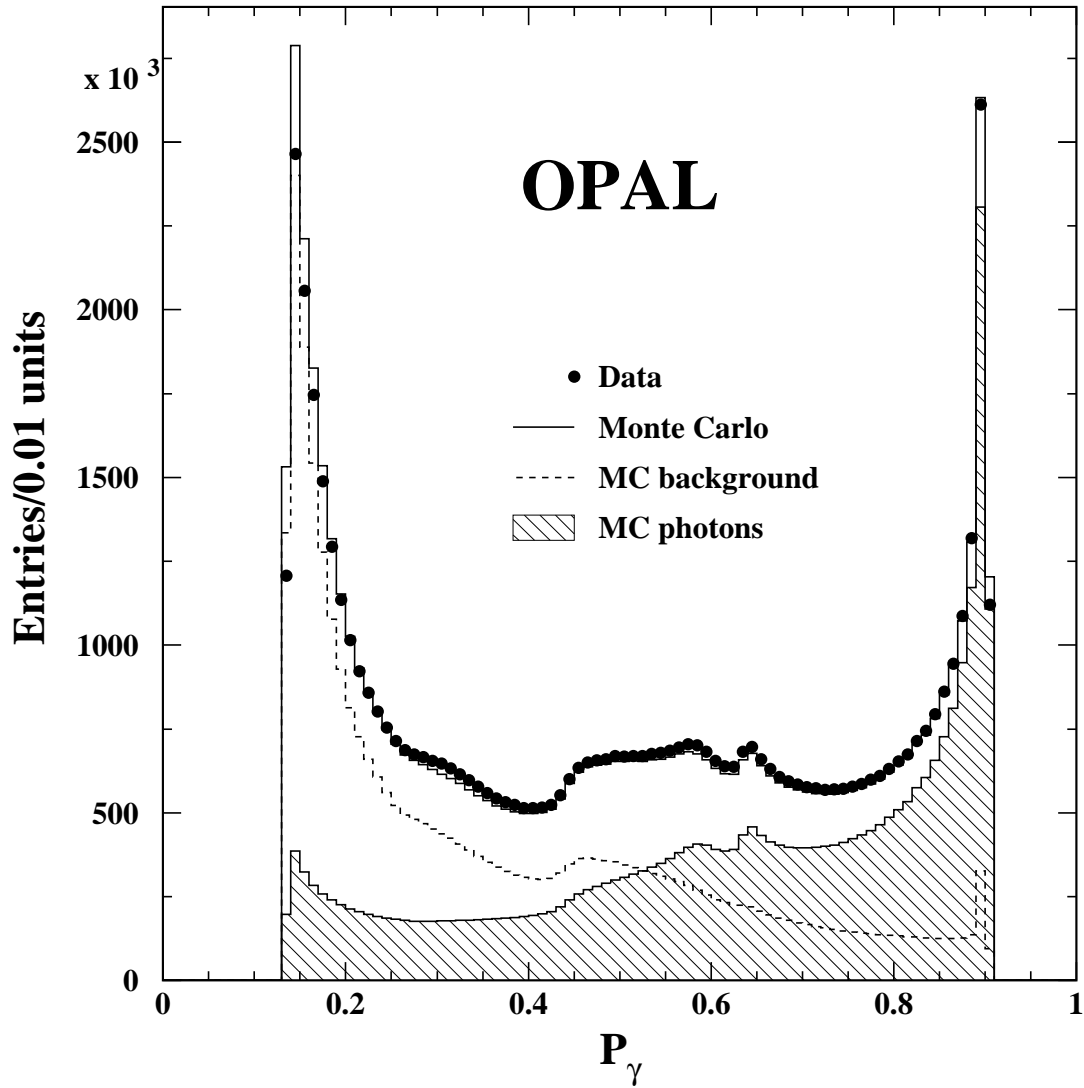
,height=17cm,bbllx=0cm,bblly=0cm,bburx=21cm,bbury=21cm}
\caption{ \label{fig-pg}
Distribution of \Pg, the variable used for the selection
of calorimeter photons in the measurement of the photon rates.
The points are data and the histogram represents
the Monte Carlo simulation.
The Monte Carlo signal (dashed histogram) and background
(hatched histogram) are also shown.
The Monte Carlo signal has been normalised to the
measured photon rate and the background to the
number of events in the data sample.}
\end{figure}

The purity of the Monte Carlo photon sample, \Pg,
has a smooth dependence on $S$ which is easily
parameterized by an analytical function.
The distribution of \Pg\  is shown in fig.~\ref{fig-pg}.
It is well reproduced by the simulation, in particular
in the region \Pg\   $>$ $0.5$, where the signal is expected
to dominate. The variable \Pg\  is used
for the systematic studies of photon samples
of varying purity described in section~\ref{sect-grate}.

\subsection{Reconstruction of photon conversions}
\label{sect-reccon}

According to the Monte Carlo simulation, 
7\% of the photons in the angular range $|\cos\theta|<0.75$
convert to \epem\   pairs in the volume of the central tracking chambers.
It is useful to compare the results obtained using this sample
with those obtained from photons detected in the calorimeter,
since they are affected by different systematic uncertainties.

The selection of photon conversions is optimised to have a high efficiency
at low momentum and a good angular resolution, and to be insensitive to
details of the detector simulation.
Conversions are observed as two oppositely charged tracks
in the central detector.
The two tracks must have impact parameters relative to the
primary event vertex greater than 300 $\mu$m in the $r-\phi$ plane,
and at least one track must
have space point measurements inside the $z$-chambers.
The average of the two points where the tracks are parallel
in the $r-\phi$ plane is taken as the point of conversion.
The pair topology is required to be consistent with the expectation
for a conversion:
\begin{itemize}
\item{the distance between the two points where the tracks are parallel
      in the $r-\phi$ plane is required to be less than 1 cm,}
\item{the radial coordinate of the point of conversion, $r$,
      must be greater than 3 cm,}
\item{the reconstructed photon must have an impact parameter relative to
      the primary event vertex in the $r-\phi$ plane smaller than 5 cm,}
\item{the absolute value of the difference between $r$ and the
      radial coordinate of the first space point measurement on
      either track must be less than 20 cm.}
\end{itemize}
According to the Monte Carlo, each of these topological cuts
removes less than 2\% of the conversion sample.
This loose selection is sufficient to obtain a 90\% pure sample
for $r>$ 50 cm.
For $r<$ 50 cm, the background increases because of the large
track density and it is further required that the d$E$/d$x$
measurements of the two tracks each have a probability greater than
1\% for the electron hypothesis.
According to the simulation, this cut removes 4\% of the conversions,
achieving a purity of 85\% for the entire conversion sample.
No further cuts are applied since this purity is sufficient to obtain
invariant mass spectra of pairs of photon candidates where the background
is dominated by random combinations of genuine photons.

The direction of the photon in $\theta$ is computed from the track
parameters with the added constraint that the photon comes from the
primary event vertex.
The polar angle of the photon direction is required to be in the
same fiducial region used for the calorimetric sample
($|\cos\theta|$ $<$ 0.75).

\begin{figure}[tbp]
\epsfig{file=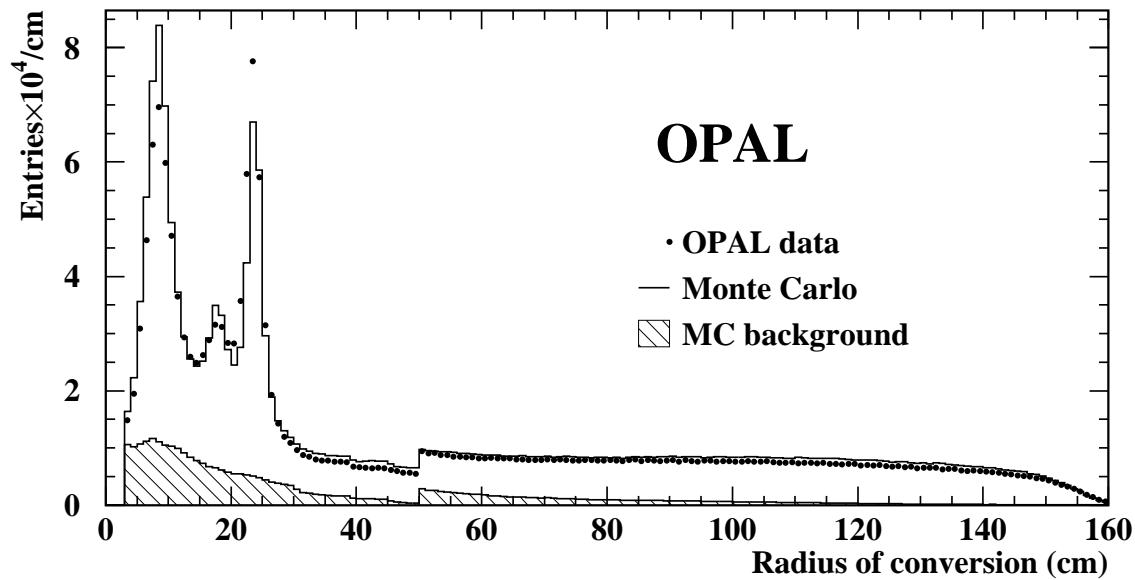
,height=10cm,bbllx=0cm,bblly=9cm,bburx=21cm,bbury=21cm}
\caption  { \label{fig-convrad}
Radial coordinate $r$ of the photon conversion, in the data
(dots) and in the Monte Carlo (histogram). The shaded
histogram shows the contribution from background, according
to the Monte Carlo.
The Monte Carlo sample is normalised to
the same number of events.
The peaks close to 9, 18 and 23 cm correspond to
concentration of material in the detector.
The abrupt cut at $r$ = 50 cm is due to an
additional cut on the d$E$/d$x$ of the two electron tracks.}
\end{figure}

The distribution of the radial coordinate $r$ of the conversion points
is shown in figure~\ref{fig-convrad}.
The Monte Carlo simulation reproduces the overall shape due to the
local concentrations of the material in the detector.
However, the numbers are not reproduced exactly:
for example, in the data 58.0\% of the conversions lie below $r=50$ cm,
while in the simulation this fraction is 59.2\%.
Such discrepancies are considered in the following estimation of the
systematic errors.

According to the simulation, in the energy range from 0.15 to 20 GeV,
the angular resolution on the direction of the photon conversions
decreases from 10 to 1 mrad in $\phi$ and from 24 to 15 mrad
in $\theta$, and the energy resolution is approximately constant
at 4\%.
Approximately two thirds of the photon conversions in the fiducial region
are reconstructed and selected, corresponding to an average photon efficiency
of about 3\%.

\subsection{Evaluation of the photon yield}
\label{sect-grate}

As the sizes of the photon candidate samples are large,
the precision of the measured yields is expected to
be limited by systematic uncertainties.
It is therefore important to compare the yields derived
from samples obtained with different selection procedures,
using different Monte Carlo predictions for the photon
efficiencies and different methods to subtract the background.
Here, the size of the final systematic errors
is reduced by incorporating these tests in the determination of
the yield itself.

In a first step, the number $n(E_{\gamma},\Delta E_{\gamma})$
of photons per hadronic \Zzero\  decay in an energy bin of width
$\Delta E_{\gamma}$ centred at $E_{\gamma}$ is derived using
five different data samples (noted by the index $i$),
three Monte Carlo samples (index $j$) and
two methods to estimate the background (index $k$):
\begin{eqnarray} \label{eqn-rate}
n^{i,j,k}(E_{\gamma},\Delta E_{\gamma}) = \frac{
   N^{i}_{\mbox{\scriptsize{candidates}}}(E_{\gamma},\Delta E_{\gamma}) -
   N^{i,j,k}_{\mbox{\scriptsize{bkg}}}(E_{\gamma},\Delta E_{\gamma}) }
                    { \epsilon^{i,j}(E_{\gamma},\Delta E_{\gamma})
                      N_{\mbox{\scriptsize{Z}$^0$}} } \mbox{ . }
\end{eqnarray}
Here $N^{i}_{\mbox{\scriptsize{candidates}}}$
is the total number of photon candidates in the data,
$N^{i,j,k}_{\mbox{\scriptsize{bkg}}}$
is the predicted number of fake photons,
$\epsilon^{i,j}$ is the efficiency for
photons in that energy bin, and
$N_{\mbox{\scriptsize{Z}$^0$}}$ is the number of \Zzero\  decays.
In the Monte Carlo, the background is defined as those candidates that
are not unambiguously associated to a single photon, or photons coming
from bremsstrahlung radiation or decays of particles produced in
interactions with the material of the detector.
For example, the Monte Carlo predicts that about half of the candidates
above 15 GeV result from the overlap of the two photons from high-energy
\piz\  decays. These unresolved photons cannot be counted appropriately
and are therefore considered as background.

The five data samples are the conversion sample, three calorimetric
samples
with requirements \Pg\   $>$ 0.0, 0.5 and 0.75, and the calorimetric
sample with each entry weighted by \Pg.
The three Monte Carlo samples used for evaluating
$N^{i,j,k}_{\mbox{\scriptsize{bkg}}}$ and $\epsilon^{i,j}$
are based on the JETSET tunes of ref.~\cite{bib-tunenew}
and~\cite{bib-tuneold}, and the HERWIG tune of ref.~\cite{bib-tunenew}.
The two prescriptions for the determination of
$N^{i,j,k}_{\mbox{\scriptsize{bkg}}}$ are to take the Monte Carlo
prediction and to scale it either according to the number of events
or the number of photons.
In total, 30 energy-dependent yields $n^{i,j,k}(E_{\gamma},\Delta E_{\gamma})$
are obtained.

In a second step, the central value for $n(E_{\gamma},\Delta E_{\gamma})$
is obtained from a weighted average of the 30 analyses.
The average is first performed on the background assumptions $k$,
using the quadratic sums of the statistical errors on the data and on
the Monte Carlo as weights.
This results in averaged yields $n^{i,j}$ and the rms deviation
around this mean, $\sigma^{i,j}_{bkg}$, is taken as
the systematic error on the choice of background assumptions.
This error is added in quadrature to the total error,
and the $n^{i,j}$'s are averaged over the choice of Monte Carlo $j$,
using the new total error as a weight.
This in turn yields new averages, $n^i$, and the error associated with
the choice of Monte Carlo $\sigma^i_{MC}$, which are again
added in quadrature to the total error.
Next, the same procedure is applied to the four calorimetric samples,
resulting in an average for these samples, $n^{cal}$, and
a systematic error associated with the use of \Pg.
In the end, two independent measurements of
$n(E_{\gamma},\Delta E_{\gamma})$ are obtained, one
from the calorimetric sample, and one from the conversion sample.
At the same time, the three systematic errors associated to the
variations of $i$, $j$ and $k$ have been calculated.
The weights used to evaluate the average yields are also used
to calculate the average error associated to each source,
assuming that the samples being combined are fully correlated.
For simplicity, this conservative assumption is also applied to the
statistical errors on the data and Monte Carlo samples.
This is justified because the statistical errors represent
a small fraction of the total errors and, in general,
there is a large overlap between the samples being combined.
A notable exception is the combination of the calorimetric
and conversion samples, which is discussed in section~\ref{sect-combi}.
%
%The comparison and combination of the calorimetric and conversion
%results are discussed in section~\ref{sect-combi}.

The averaging procedure is performed separately for each energy
bin in order to take into account the variation with energy of the
nature and the size of the systematic uncertainties.
As a check, the order of the averaging procedure is reversed,
and the resulting yields and errors are compared.
No significant differences are observed, indicating that
the different systematic tests are largely uncorrelated.
However, the variations arising from a given test are 
assumed to be fully correlated from one energy bin to another.

Two corrections are applied to the average yields.
The first one is for the difference in energy calibration
between the data and the Monte Carlo explained
in detail in section~\ref{sect-pizpeak}.
The second accounts for the difference between the data
and the Monte Carlo
in the number of photons initiating a shower before
reaching the calorimeter.
A study of the fraction of calorimeter photons with an
associated signal in either the presampler or time-of-flight
reveals that the Monte Carlo underestimates by 2\% the probability
of initiating a shower before reaching the calorimeter.
The efficiency is corrected for this effect.
It is
important only for photons with an energy comparable to the average
energy lost before reaching the calorimeter, i.e., about 200 MeV.

The numbers of photons per hadronic \Zzero\  decay in the
energy range \xE\  $>$ 0.003 obtained from the calorimetric ($\gamma$)
and conversion ($\gamma_c$) measurements are given
in table~\ref{tab-gamerr} together with the values
of each systematic uncertainty.
The sources of these uncertainties are:
\begin{itemize}
\item{The statistical error on the Monte Carlo samples used to
      calculate the efficiency.}
\item{The variations observed using different Monte Carlo samples,
      obtained from the averaging procedure.}
\item{The error associated with \Pg,
      obtained from the averaging procedure.}
\item{The variations observed when using the different background
      assumptions,
      obtained from the averaging procedure.}
\item{Some of the background comes from photons
      produced in interactions with the material of the detector.
      There are indications that this effect is not exactly reproduced
      by the Monte Carlo; see for example ref.~\cite{bib-opalpkpi} and the
      discussion on the electron bremsstrahlung in
      section~\ref{sect-pizpeak}.
      Therefore, half of the Monte Carlo prediction for
      this background is conservatively taken as an uncertainty.}
\item{As the measured yields depend on the exact energy calibration,
      the analyses are repeated by shifting the energy scale by 1\%
      and the difference in the rates is taken as a systematic error.
      The size of this shift is justified by the calibration studies
      described in the following section on \piz\
      and $\eta$ reconstruction.}
\end{itemize}

The variation of the cuts on \Pg\  does not cover
all sources of systematic uncertainties on the quality of the simulation.
The following systematic errors are considered:
\begin{itemize}
\item{For the calorimeter data, the error labelled {\em simulation}
      in table~\ref{tab-gamerr} is the quadratic sum of
      the uncertainty on the correction
      of the probability of initiating a shower before reaching
      the calorimeter and the yield variation observed when
      the criteria for the association of charged
      tracks to energy deposits are varied.}
\item{For the conversion data, the  error labelled {\em simulation}
      is obtained by removing each of the selection cuts in turn.
      The differences between the number of accepted tracks in the data
      and the Monte Carlo obtained when each cut is removed are added
      in quadrature. }
\end{itemize}

For the calorimetric measurement,
the largest source of systematic uncertainty 
is the difference between the efficiencies derived
from the three simulations (table~\ref{tab-gamerr}).
In contrast to ref.~\cite{bib-lpiz}, where the most important
error arises from the choice of using either JETSET or HERWIG
for the determination of the efficiencies, here the largest
difference is observed between the two samples generated
using the JETSET versions tuned in refs.~\cite{bib-tunenew}
and~\cite{bib-tuneold}.
The efficiencies obtained with the HERWIG 5.9 sample
are consistent those obtained with the first JETSET sample.
Notable differences between the two JETSET samples are the
inclusion of $L=1$ mesons~\cite{bib-tunenew} and consequent
changes to the $\omega$ and $\eta'$ rates by factors of 1.5 and 4.5,
respectively.
These differences affect the photon efficiencies due to
the presence of neighbouring particles.
Another difference between the two samples is the version
of the detector simulation program~\cite{bib-gopal}.

For the conversions, the largest source of uncertainty is
that associated with the simulation of the selection cuts
(table~\ref{tab-gamerr}).
The error is 6.8\% over most of the energy range and is
dominated by effects related to the inadequacy in describing
the distribution of the radial coordinate of the conversions.
At the lowest energy it increases to 10\% due to uncertainties
in reconstructing tracks with small curvature radii.

As a consistency check, the photon yields obtained from
the calorimetric and the conversion sample have been compared
in 10 data sub-samples corresponding to different data-taking
periods, spanning six years of operation of OPAL at \Zzero\
energies.
The calorimeter efficiency varies by less than 1\% and the
number of conversions varies with an rms of 2.6\%,
well within the systematic errors estimated for those channels.

The differential cross sections of inclusive photon production
as a function of \xE\   and \lnxp\  
are presented, interpreted and discussed
together with those for the light mesons in section~\ref{sect-results}. 

%%%%%%%%%%%%%%%%%%%%%%%%%%%%%%%%%%%%%%%%%%%%%%%%%%%%%%%%%%%%%%%%%%%%%%%%%

\section{ The channels {\boldmath $\pi^0\rightarrow\gamma\gamma$} and
          {\boldmath $\eta\rightarrow\gamma\gamma$} }
\label{sect-gg}

In this section, \piz\  and $\eta$ mesons are reconstructed as pairs
of photons. 
The branching ratios of the decays \pigg\  and \etagg\
are (98.80 $\pm$ 0.03)\% and (39.25 $\pm$ 0.31)\%,
respectively~\cite{bib-pdg}.
The numbers of \piz\ and $\eta$ mesons in the data and Monte Carlo
samples are determined from fits to the invariant mass spectra
of the photon pairs.
The selection of candidates is presented, followed by a description
of the fits to the invariant mass spectra and the determination of
the meson yields.
As in the photon case, the yields and some systematic errors
are obtained by averaging
results based on different samples and using various analysis
methods.

\subsection{ {\boldmath $\pi^0\rightarrow\gamma\gamma$} and
             {\boldmath $\eta\rightarrow\gamma\gamma$} selection   }
\label{sect-recgg}

The \piz\  and $\eta$ candidates are obtained by combining in turn
all pairs of photon candidates.
In the $\eta$ selection, the energy of each photon is required to
be larger than 0.3 GeV.
At this stage, the combinatorial background is large,
and it is not possible to extract the \piz\
and $\eta$ yields from the invariant mass distribution
of the photon pairs.

To reduce the background, the
probability \Ppi\  that a photon pair comes from a \pigg\ 
decay as a function of a set of input variables is estimated
using the same method as for \Pg\  in section~\ref{sect-recgam}.
The input variables for each photon are those shown in fig.~\ref{fig-gvar}a,
b, d, f, and g, together with the opening angle $\theta_{ij}$
of the photon pair, the number of additional photons in
cones of opening angle $\theta_{ij}$ around each of the two photons
and the helicity angle of the photon calculated in the
$\gamma\gamma$ rest frame.
The invariant mass of the pair is excluded from the input variables
so that it can be used later on for extracting the yields.
The same method is applied to the decays involving
conversions\footnote{%
For the conversions, the input variables corresponding
to fig.~\ref{fig-gvar}b, d, f, and g are not relevant and
are therefore not used.}
and the three functions
$P_{\gamma\gamma}^{\mbox{\scriptsize{\pigg}}}$,
$P_{\gamma\gamma}^{\mbox{\scriptsize{\piggc}}}$ and
$P_{\gamma\gamma}^{\mbox{\scriptsize{\pigcgc}}}$
are determined separately.
Similarly, the equivalent functions for the $\eta$,
$P_{\gamma\gamma}^{\mbox{\scriptsize{\etagg}}}$, 
$P_{\gamma\gamma}^{\mbox{\scriptsize{\etaggc}}}$ and
$P_{\gamma\gamma}^{\mbox{\scriptsize{\etagcgc}}}$,
are evaluated separately.
The purity of the \piz\  Monte Carlo sample is evaluated using
a $\Delta m$=0.1 GeV/$c^2$-wide invariant mass window centred at
$m_0=$0.135 GeV/$c^2$.
For the $\eta$, the purity of the Monte Carlo sample is evaluated
using $\Delta m$=0.2 GeV/$c^2$ and $m_0$=0.5475 GeV/$c^2$.

The probability that a pair of photons $i,j$ is part of the
\piz\  or $\eta $ signal
depends not only on the \Ppi\  value for this pair,
but also on the signal probability for any other pair involving
either $i$ or $j$.
Assuming that \Ppi\  is indeed equal to the signal probability,
and assuming that the \Ppi\  values of all photon pairs in an event
are uncorrelated,
the probability that two photons $i$ and $j$ are not related to
any other photon in the event is:
\begin{eqnarray}\label{eqn-pprod}
\Pslash\   & = &
  \prod_{k \neq i,j} [ 1 - G_{ik}(m_{ik}) ] \times
  \prod_{l \neq i,j} [ 1 - G_{jl}(m_{jl}) ] \mbox{ . }
\end{eqnarray}
Here $m_{ik}$ is the invariant mass of the pair of photons $i$ and $k$,
and
\begin{eqnarray} \label{eqn-ptpi}
 G_{ik}(m_{ik}) & = & \frac{ P_{\gamma_i\gamma_k} }
     { P_{\gamma_i\gamma_k}  + ( 1 - P_{\gamma_i\gamma_k}  )
   \frac{\sqrt{2\pi}\sigma}{\Delta m}
        \exp(\frac{1}{2}(\frac{m_{ik}-m_0}{\sigma})^2)  }
\end{eqnarray}
is the \Ppi\  variable modified to take into account the mass
dependence of the purity of the \piz\  and $\eta$ signals, 
assuming that the invariant mass peaks are normal
distributions of width $\sigma$, which are determined from
the Monte Carlo.
The products in eq.~\ref{eqn-pprod} are performed over all pairs
retained by either the \piz\   or the $\eta$ selection.
If a pair passes both selections, the larger of the \piz\
or $\eta$ signal probability is retained.
The combined probability that the photons $i$ and $j$ are from the
same \piz\  or $\eta$ decay and that they do not take part in
any other decay is taken as \iPtpi\  = \iPpi$\times$\Pslash.
According to the simulations, cutting on \Ptpi\  instead of \Ppi\ 
reduces the combinatorial background by approximately 10\% in the
case of the \piz\   and by as much as 30\% in the case of the $\eta$.

Since the \piz\  and $\eta$ yields are determined from
fits to the invariant mass spectra, the effect of a cut on  \Ptpi\ 
on the shape of the invariant mass distributions has to be studied.
The value of \iPtpi\  has a monotonic correlation with the invariant
mass of the pair $m_{ij}$, which arises primarily from its dependence 
on the opening angle $\theta_{ij}$ and the number of additional photons
in the cones of opening angle $\theta_{ij}$ around $i$ and $j$.
The shifts in the position of the \piz\   and $\eta$ mass peaks
induced by cuts on \Ptpi\  are of the order of a few MeV/$c^2$
and are well reproduced by the Monte Carlo.
More importantly, it has been verified in the Monte Carlo
that the cuts on \Ptpi\   do not produce a fake \piz\  or $\eta$ peak
in the invariant mass distribution of the combinatorial background.

\begin{figure}[tbp]
\epsfig{file=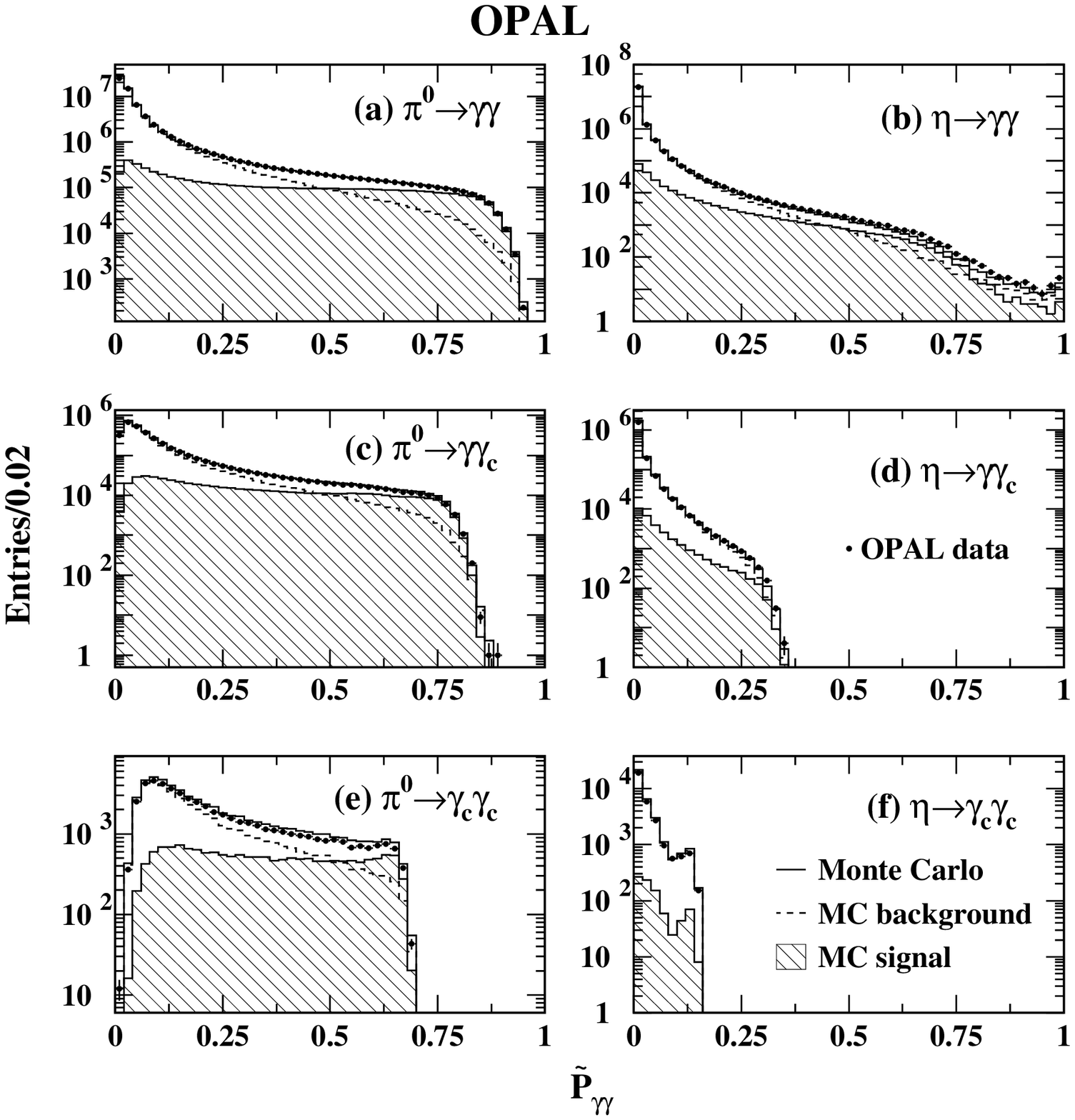
,height=17cm,bbllx=0cm,bblly=0cm,bburx=21cm,bbury=21cm}
\caption  { \label{fig-pgg} 
a-f: Distribution of \Ptpi\
for the channels 
\pigg, \etagg, \piggc, \etaggc, \pigcgc\   and \etagcgc.
The points represent the data and the histograms
represent the Monte Carlo simulation normalised to the
same number of events.
The Monte Carlo signals and backgrounds are shown
separately
as hatched and dashed histograms, respectively.}
\end{figure}

\begin{figure}[tbp]
\epsfig{file=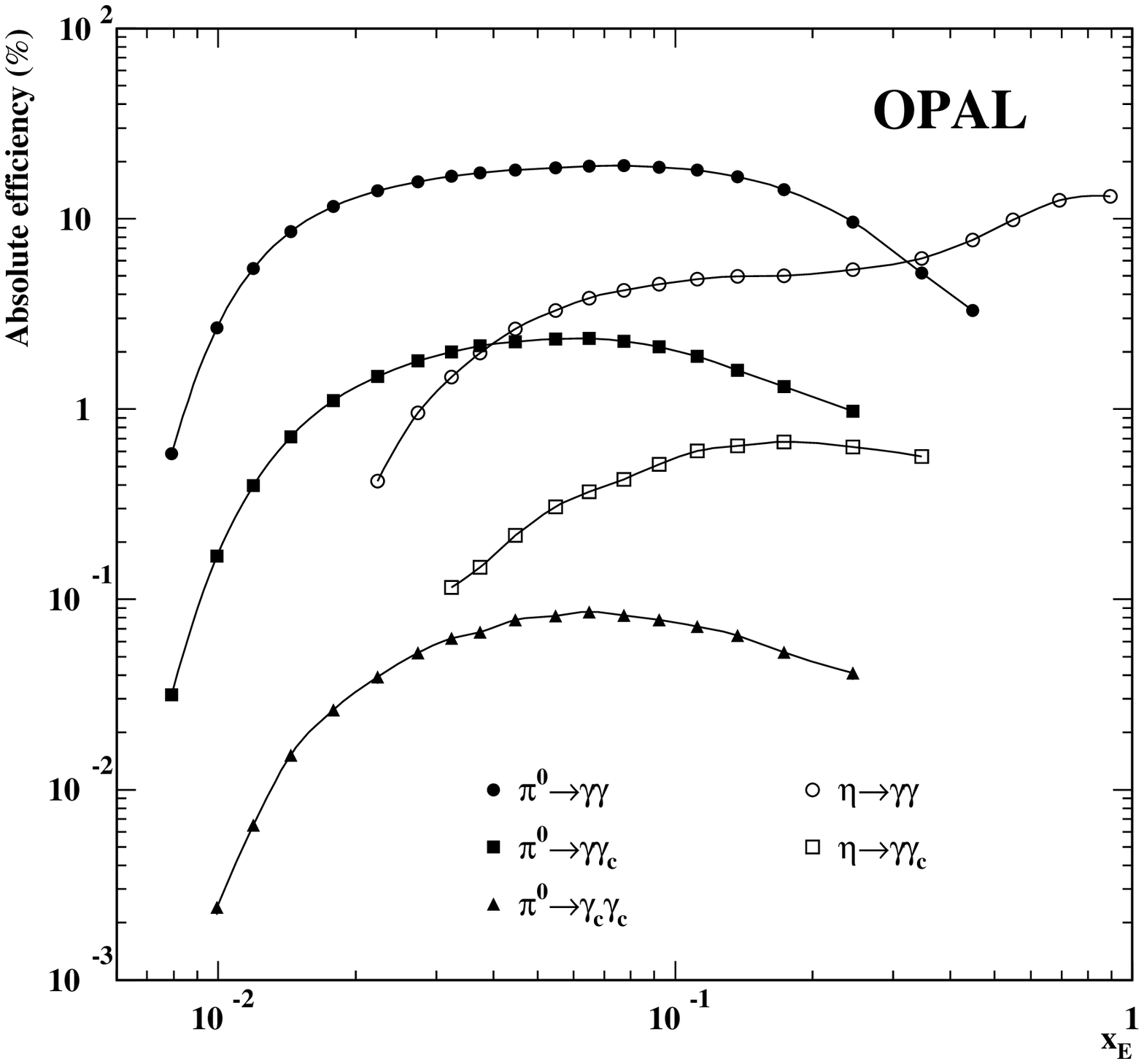
,height=17cm,bbllx=0cm,bblly=0cm,bburx=21cm,bbury=21cm}
\caption{ \label{fig-eff}
Absolute efficiency as a function of the scaled meson
energy \xE\   for the reconstruction and
identification of \piz\   and $\eta$ mesons,
according to the Monte Carlo simulation.
The efficiencies for the decay channels \pigg, \piggc, \pigcgc,
\etagg\   and \etaggc\   are shown separately.
A cut on \Ptpi$>0.1$ is applied for the \piz,
and \Ptpi$>0.05$ for the $\eta$ (see text).}
\end{figure}

Fig.~\ref{fig-pgg} shows the distributions of \Ptpi\  for all channels
and indicates that
their shapes are well reproduced by the Monte Carlo.
A more detailed discussion of possible differences will follow
in section~\ref{sect-rategg}. 
The Monte Carlo predictions for the efficiency for detecting
\piz\  and $\eta$ mesons using cuts of 0.1 and 0.05 on \Ptpi,
respectively, are shown in fig.~\ref{fig-eff}.
The efficiencies include all effects, including the known branching
ratios~\cite{bib-pdg}, so that the yields obtained from the 
$\gamma\gamma$, $\gamma\gamma_c$ and $\gamma_c\gamma_c$ samples can be
directly compared.

\subsection{Analysis of the invariant mass spectra of photon pairs}
\label{sect-pizshape}

As in the photon analysis, the determination of the \piz\  and
$\eta$ yields is repeated using different data samples and
analysis procedures, and the results are averaged to obtain
the central values.
Therefore, the fits to the invariant mass spectra are performed
separately for the  \pigg, \piggc, \pigcgc, \etagg\  and \etaggc\
samples.
The channel \etagcgc\   is not used because of low statistics.
The fit is systematically repeated using five values of the cut on \Ptpi,
two parameterizations of the background and
two parameterizations of the signal peaks.
In addition, three Monte Carlo samples are used for the determination
of the \piz\   and $\eta$ efficiencies and, for one of the background
parameterization methods, the background shape.
These variations are described below.
They amount to a total of 60 different methods to extract the
rates for each channel.

\subsubsection{Cut variations}

The cut on \Ptpi\  is varied among the values 0.1, 0.2, 0.3, 0.4
and 0.5 for the \piz\  analysis, and 0.05, 0.10,
0.20, 0.30 and 0.40 for the $\eta$ analysis.
The variations of the \Ptpi\  cut are chosen so as to result in a change
in acceptance of at least a factor two, in order to test how well \Ptpim\
is simulated by the Monte Carlo, and also to provide a wide variety of
background shapes and levels.

\subsubsection{Background parameterization}

\begin{figure}[tbp]
\epsfig{file=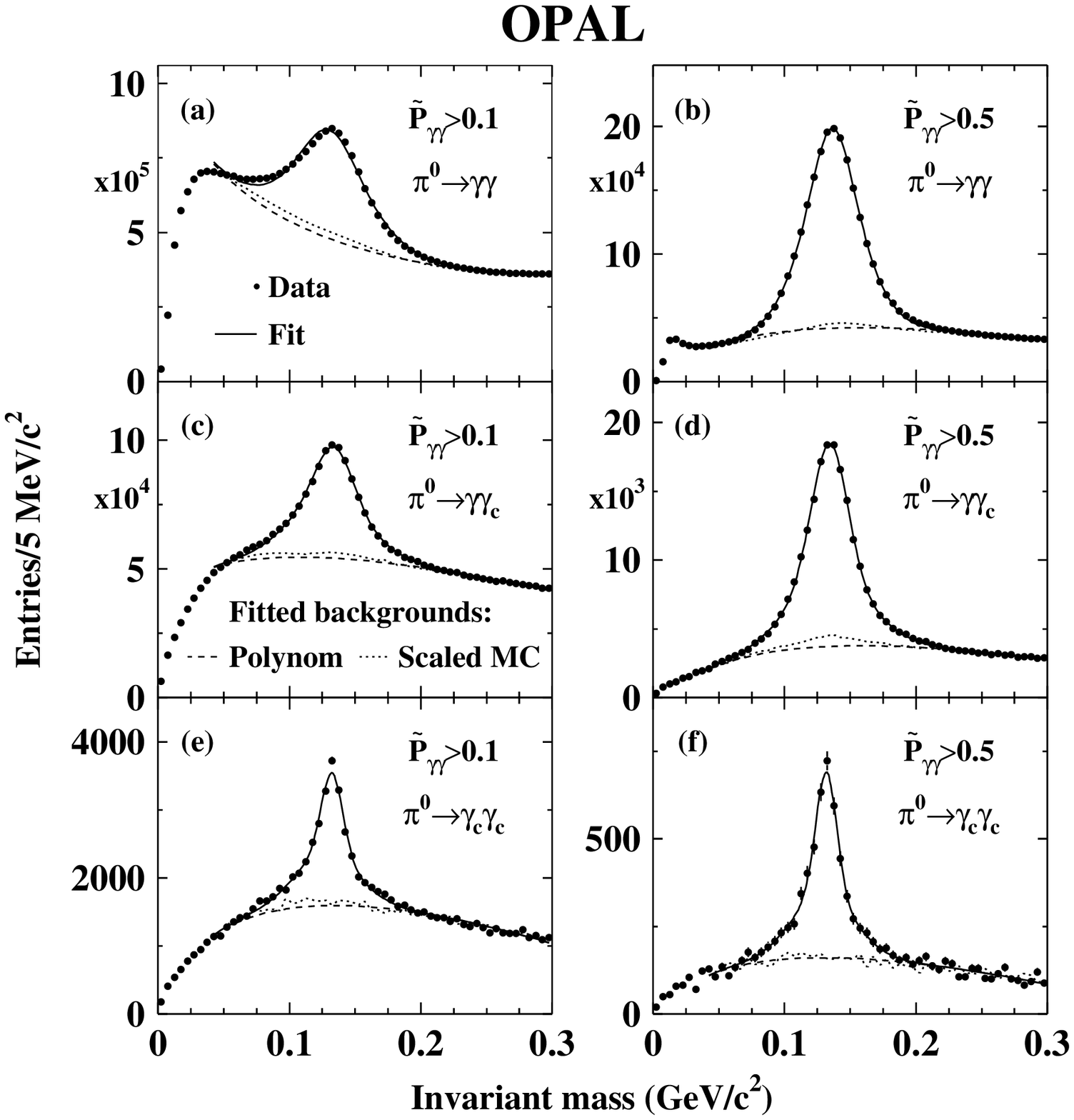
,height=17cm,bbllx=0cm,bblly=0cm,bburx=21cm,bbury=21cm}
\caption{ \label{fig-fitpi}
a-f) Fits to the invariant mass distribution of pairs of
photons, for the decay channels \pigg, \piggc\
and \pigcgc\  for the lowest and highest \Ptpi\  cut
used in the present analysis.
The points represent the OPAL data and the full
lines the fits to the data where the background (dashed lines)
is parameterized as a second-order polynomial.
The dotted lines correspond to the background evaluated
using the Monte Carlo distributions,
which may contain a small fraction of the signal in cases
where the association between photon candidates and true
photons is ambiguous (see text).}
\end{figure}

\begin{figure}[tbp]
\epsfig{file=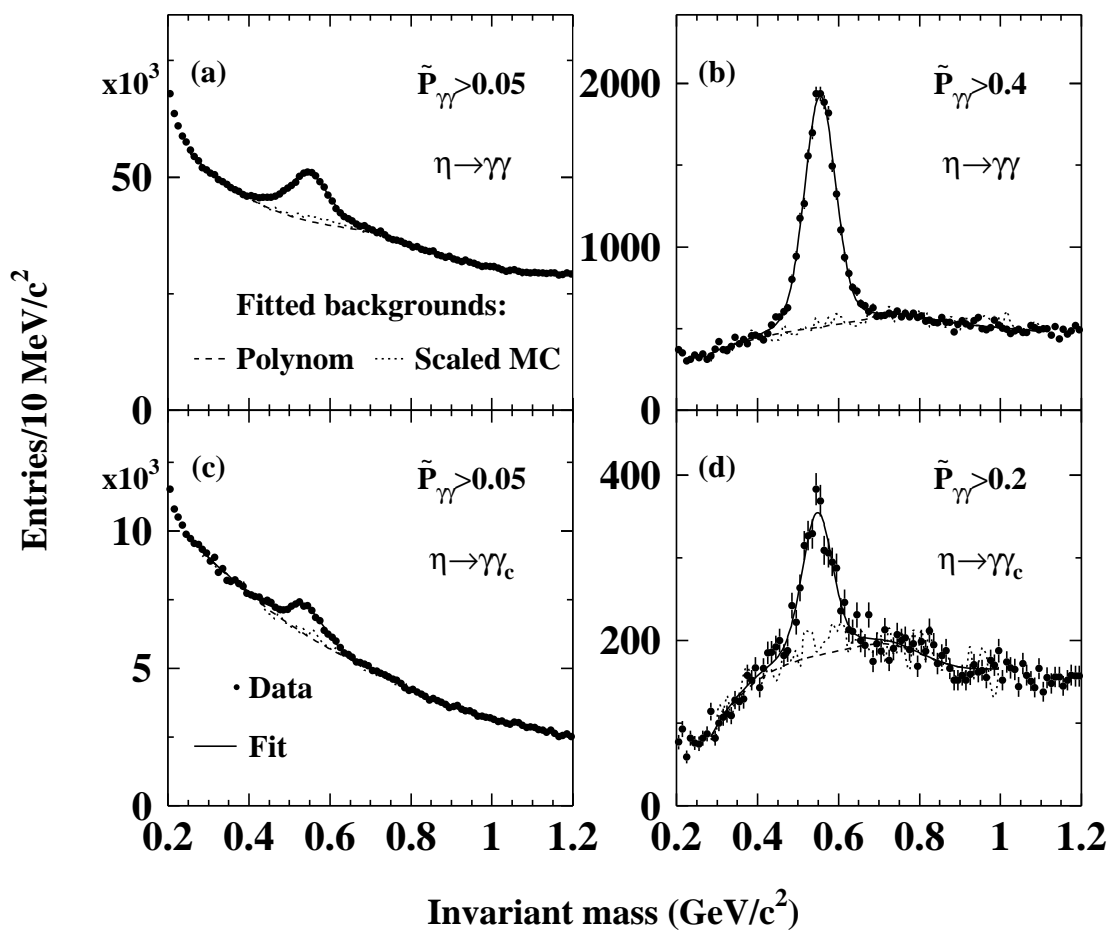
,height=17cm,bbllx=0cm,bblly=0cm,bburx=21cm,bbury=21cm}
\caption{ \label{fig-fiteta}
a-d) Fits to the invariant mass distribution of pairs of
photons, for the decay channels \pigg\  and \etaggc\
for the lowest and highest \Ptpi\  cut
used in the present analysis.
The points represent the OPAL data and the full
lines the fits to the data where the background (dashed lines)
is parameterized as a second-order polynomial plus
a Gaussian for the reflection from
$\omega\rightarrow\gamma\pi^0$ decays.
The dotted lines correspond to the background evaluated
using the Monte Carlo distributions,
which may contain a small fraction of the signal in cases
where the association between photon candidates and true
photons is ambiguous (see text).}
\end{figure}

Figs.~\ref{fig-fitpi} and~\ref{fig-fiteta}
show the invariant mass spectra for all five channels.
The two background estimations are also shown.
In the first one, the background shape is taken from a simulation
and is normalised to the number of counts outside the signal region.
Possible differences between the shape in the data and in the Monte
Carlo are taken into account by adding a linear background to the fit.
In the second method, the background is fitted using
a second-order polynomial.
An additional source of background is considered in the
region just above the $\eta$ peak.
The reflection from $\omega\rightarrow\gamma\pi^0$ decays
are taken into account using
a Gaussian with a normalisation allowed to vary
and a mass and width of 730 and 80 MeV/$c^2$, respectively.
This peak is caused by the kinematic correlation between one of
the photons from the \piz\  and the direct photon from the $\omega$,
and cannot be absorbed by the polynomial background.

The two background parameterizations are complementary;
the first one takes into account all the features of the
background shape predicted by the simulations, while the
second does not depend on the details of the Monte Carlo.
The area of the signal peaks obtained with the two methods
are in general not identical, as shown in figs.~\ref{fig-fitpi}
and~\ref{fig-fiteta}.
However both methods should give the same yields if the
efficiency is determined by applying the same parameterization
to the Monte Carlo.
As a result, the errors due to the background parameterization are smaller
than is suggested by the difference between the two background
estimates in figs.~\ref{fig-fitpi} and~\ref{fig-fiteta}.
Part of this difference comes from an ambiguity in the definition of
the background in the Monte Carlo.
In a few percent of the cases, a shower reconstructed in the calorimeter
has contributions from more than one incident particle and cannot be clearly
associated to any one of them.
However, the contribution of a photon to this shower might be important enough
that when it is combined with the photon coming from the same \piz\  decay,
the invariant mass of the pair may be very close to the mass of the \piz.
Even though such pairs may produce a small \piz\  peak, the first
method consider them as being part of the background while the second
will tend to treat them as part of the signal.
The averaging procedure takes into account this ambiguity and the related
uncertainty.

\subsubsection{Signal parameterization}
\label{sect-pizpeak}

\begin{figure}[tbp]
\epsfig{file=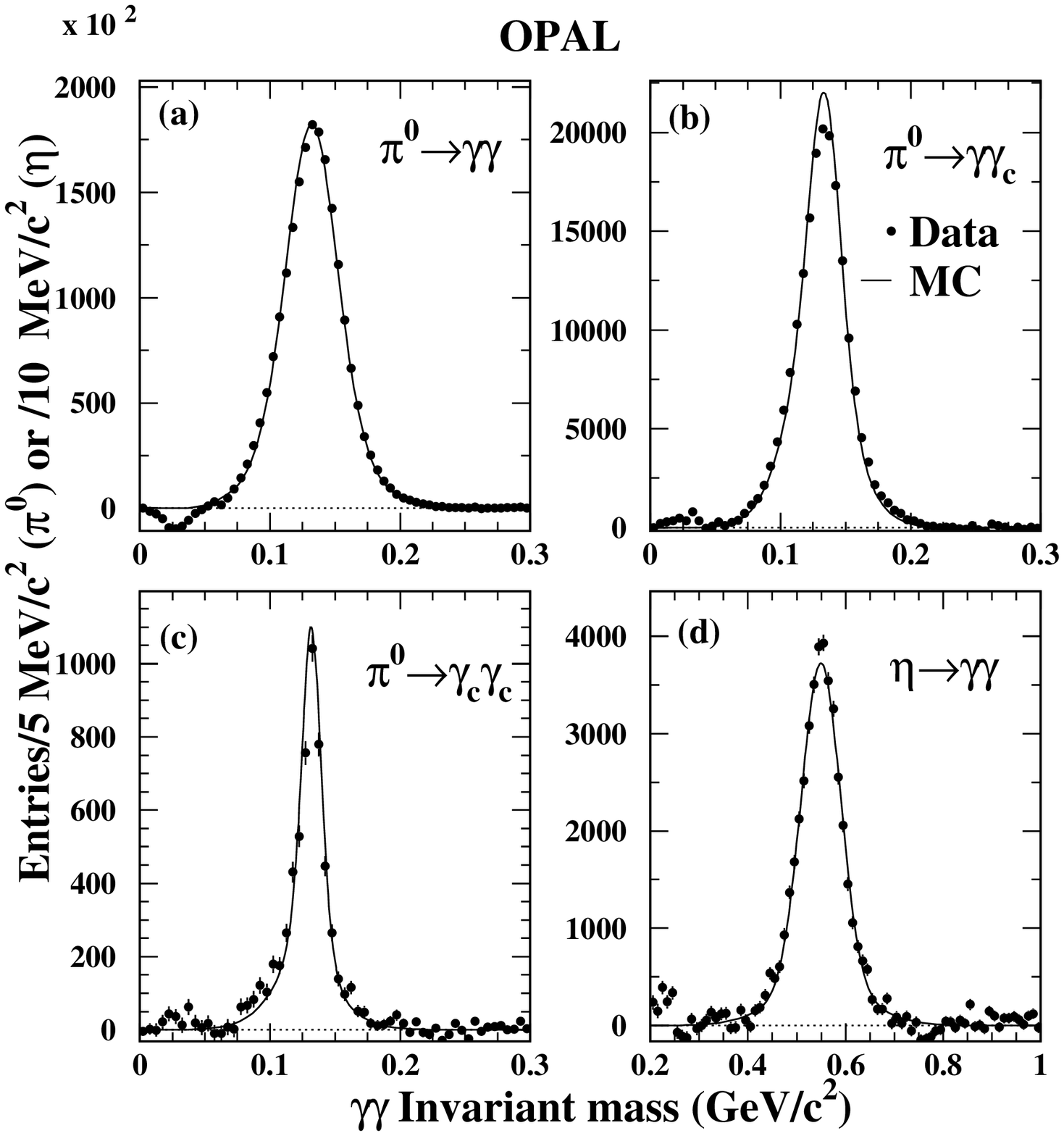
,height=17cm,bbllx=0cm,bblly=0cm,bburx=21cm,bbury=21cm}
\caption{ \label{fig-shape}
a-d) Mass peak for the decay channels \pigg, \piggc, \pigcgc\
and \etagg. The distributions are obtained by subtracting
the fitted background shape from the raw spectra.
The points represent the OPAL data and the full
curves the fit to the Monte Carlo.
The Monte Carlo distributions are normalised to 
the area in the data.
A cut on \Ptpi$>0.3$ is applied for the \piz,
and \Ptpi$>0.2$ for the $\eta$ (see text).}
\end{figure}

Fig.~\ref{fig-shape} shows  the shape of the mass peak
for the samples \pigg, \piggc, \pigcgc\   and \etagg\
in the data and in the simulation.
In all cases the distributions are obtained by subtracting
the fitted background from the raw spectrum.
In the fits, the shapes are alternatively parameterized either as
a Gaussian function,
or as the more complex functions described below.

With high statistics samples such as those in fig.~\ref{fig-shape},
the non-Gaussian structure of the signal peaks is apparent.
The peaks for the \pigg\  and \etagg\  decays are
each described by two Gaussians centred at the same mass.
The centroid, widths and amplitudes of the two Gaussians
are determined independently for the data and the Monte Carlo.
The centroids and widths are well described by the simulation
(fig.~\ref{fig-shape}a and d).
In the case of \piggc\  and \pigcgc\  decays
(fig.~\ref{fig-shape}b and c),
bremsstrahlung from the conversion electrons produces a
pronounced tail toward low invariant masses~\cite{bib-convtail}.
In the fit, this tail is described by an exponential convoluted with
the fitted mass resolution.
The amplitude and decay constant are determined independently for
the data and the Monte Carlo.
In the case of the \etaggc\  signal,
the tail is neglected because of the low statistics.

The comparison of the centroids of the \piz\  and $\eta$ signals
in the data with those obtained in the simulations provides
an important check of the energy calibration.
The corrections for all five channels are of order 1\%
over most of the energy range. This sets the scale for
the uncertainty on the energy calibration.

The quality of the detector simulation and its impact on the
analysis can be assessed by comparing the widths and the
tails of the mass peaks in the data and in the Monte Carlo.
The Monte Carlo reproduces well the peak shapes (fig.~\ref{fig-shape}),
except for the tail toward low masses for combinations
involving one or two low-energy conversions.
Such tails have been noticed before~\cite{bib-convtail} and
they are due mostly to bremsstrahlung of the conversion electrons.
The Monte Carlo simulates correctly the slope of these tails,
but underestimates their amplitude by as much as a factor
of two.
The analysis should not be affected by an inadequate modelling of
these tails, since they are free parameters of the fit.
In addition, the systematic error includes the effect of
neglecting them by assuming Gaussian peak shapes,
and the effect is further tested in section~\ref{sect-combi}
by comparing the results derived from the $\gamma$ and
$\gamma_c$ samples, and from the $\gamma\gamma$, $\gamma\gamma_c$ and
$\gamma_c\gamma_c$ samples.

\subsubsection{Monte Carlo simulations}

The three Monte Carlo samples described in section~\ref{sect-grate}
are alternatively used to evaluate the \piz\  and $\eta$ efficiencies.
When the first background parameterization method is used,
the shape of background is taken from the corresponding simulation
for consistency.

\subsection{Determination of the \piz\  and $\eta$ yields}
\label{sect-rategg}

The \piz\  and $\eta$ yields and their errors are determined separately
for the channels \pigg, \piggc, \pigcgc, \etagg\  and \etaggc\   using
the averaging method described in section~\ref{sect-grate}.
As the numbers of mesons $N^{i}_{\mbox{\scriptsize{candidates}}}$
are obtained from fits to the invariant mass spectra,
the background $N_{\mbox{\scriptsize{bkg}}}$ entering eq.~\ref{eqn-rate}
is the contribution from \piz\   produced in interactions
with the material of the detector, which is taken from the simulation.
As in section~\ref{sect-grate}, the yields are corrected for the
difference between the Monte Carlo and the data in the energy calibration
and in the probability that a photon initiates a shower before reaching
the calorimeter.

The numbers of \piz\  and $\eta$ mesons per hadronic \Zzero\  decay
in the energy ranges covered by the present measurement
are given in tables~\ref{tab-pizerr} and~\ref{tab-etaerr}
together with the values
of each of the systematic uncertainties.
These are:
\begin{itemize}
\item{The statistical error on the Monte Carlo samples used to
      calculate the efficiency.}
\item{The variations observed in the averaging procedure
      using different Monte Carlo samples.}
\item{The error associated with \Ptpi,
      obtained from the averaging procedure.}
\item{The variations observed in the averaging procedure
      when using the different background parameterizations.}
\item{The variations observed in the averaging procedure
      when using the different signal parameterizations.}
\item{The uncertainty associated with the \piz\  produced in
      interactions with the material of the detector, evaluated as
      in section~\ref{sect-grate}.}
\item{The uncertainty associated with the energy calibration,
      evaluated as in section~\ref{sect-grate}.}
\item{The first method to evaluate the background requires
      the definition of a mass range used for the normalisation
      of the Monte Carlo prediction to the data.
      This procedure is not exact because of the presence of
      tails in the invariant mass distributions of the signal.
      The bias on the yields resulting from the choice of the
      normalisation range is estimated using the Monte Carlo,
      and its size is taken as the systematic error associated
      with this method of evaluating the background.}
\item{The simulation uncertainties not covered by the variations
      of the cut on \Ptpi, i.e., the errors labelled {\em simulation}
      in table~\ref{tab-gamerr}, propagated to the $\gamma\gamma$,
      $\gamma\gamma_c$ and $\gamma_c\gamma_c$ samples according
      to the number of calorimetric and conversion photons in
      the pairs.}
\end{itemize}

As a consistency check, the position of the \piz\
and $\eta$ mass peaks has been measured in 10 data sub-samples
corresponding to different data-taking periods, spanning
six years of operation of OPAL at \Zzero\  energies.
The energy scale varies by less than the 1\% systematic
uncertainty ascribed to it.
The fluctuations in the extracted \piz\  and $\eta$ yields
are of the same size as those observed in section~\ref{sect-grate}
for the numbers of calorimetric and conversion photons.

The dominant systematic uncertainties on the \pigg, \piggc\
and \pigcgc\  yields (table~\ref{tab-pizerr}) are those that
also affect the corresponding $\gamma$ and $\gamma_c$
measurements (table~\ref{tab-gamerr}),
namely the variation of the efficiency in the different
Monte Carlos for the calorimetric data, and the
simulation uncertainty associated with reconstruction cuts
for the conversion.
However, the size of these uncertainties are approximately
the same as those associated with the fits to the invariant mass
distributions.

The largest systematic error for the \etagg\  channel comes from the
variation of the cut on \Ptpi\  (table~\ref{tab-etaerr}).
This is due to the difference between the shape of the  \Ptpi\
distributions in the data and in the Monte Carlo, which becomes
clearly visible above \Ptpi$>0.5$ in fig~\ref{fig-pgg}b.
Several checks on the $\eta$ sample were performed in order to
understand the origin of this discrepancy.
A possible explanation is that the Monte Carlo simulations
underestimate the number of isolated $\eta$ mesons.
This was verified with $\eta$ samples selected requiring
that there should be no other photons or charged tracks within
a cone of half-angle 15\degree\  from the $\eta$.
The simulation underestimates the number of these isolated
$\eta$ mesons with $x_E>0.1$ by a factor 2.07 $\pm$ 0.11,
while the combinatorial background is well reproduced.
The same problem affects \piz\   mesons, for which
this factor is measured to be 1.99 $\pm$ 0.05.
The factors are the same for $\gamma\gamma$ and $\gamma\gamma_c$
samples.
They are similar to those observed in ref.~\cite{bib-lisogam}.
Thus, there is clear evidence that
the Monte Carlo simulation underestimates significantly
the number of very isolated \piz\  and $\eta$ mesons.
These mesons are systematically associated with large values of \Ptpi.
Fortunately, they constitute only a few percent of the total sample of
identified mesons and their impact on the inclusive rate
is very small.
However, they are an indication of yet unexplained shortcomings
in current Monte Carlo models, and have to be taken into account
in the evaluation of the systematic error.
In the present analysis, the error associated with the variation of
\Ptpi\  corresponds to a decrease of acceptance by a factor of two
and probes the behaviour of mesons with significantly different
environments.
An additional check is performed in sect~\ref{sect-combi}
by comparing the $\gamma\gamma$ and \threepi\  channels
which have different sensitivity to variations of the cut on \Ptpi.

The differential cross sections as a function of \xE\
and \lnxp\  are presented, interpreted and discussed
together with those for the photons and the other mesons in
section~\ref{sect-results}. 

%%%%%%%%%%%%%%%%%%%%%%%%%%%%%%%%%%%%%%%%%%%%%%%%%%%%%%%%%%%%%%%%%%%%%%%%%%%%
                                                           
\section{The decay channels {\boldmath $\pi^{\pm}\pi^0$},
                            {\boldmath $\pi^{\pm}\eta$},
                            {\boldmath $\pi^+\pi^-\pi^0$} and
                            {\boldmath $\pi^+\pi^-\eta$} }
\label{sect-charged}

The reconstruction of \piz\   and $\eta$ mesons offer the
possibility to reconstruct
the dominant \rhopipi\  and \azetapi\  decays,
the $\eta,$\omegapipipi\  decays
with branching ratios of (23.2 $\pm$ 0.5)\% and
(88.8 $\pm$ 0.7)\%
and the \etapetapipi\  decay
with a branching ratio of (43.7 $\pm$ 1.5)\%\cite{bib-pdg}.
The method to evaluate the meson yields follows closely
that used in section~\ref{sect-gg}.
Namely, the numbers of mesons are obtained from fits
to the invariant mass spectra of the meson decay products
and the final results and some systematic errors are
calculated by averaging the yields obtained using different
data samples and various analysis methods.
Two notable changes with respect to section~\ref{sect-gg}
are the slight modification of the \piz\
and $\eta$ selection described below and the
adaptation of the fit procedure to the properties of
the observed signals and backgrounds 
presented in sections~\ref{sect-fitrho} to~\ref{sect-etapfit}.
                               
\label{sect-chargedsel}

The decay channels  $\pi^{\pm}\pi^0$, $\pi^{\pm}\eta$,
$\pi^+\pi^-\pi^0$ and $\pi^+\pi^-\eta$
are reconstructed by combining
the \piz\ and $\eta$ candidates
with either one charged track or two oppositely charged tracks.
The charged tracks are required to have at least
40 measured space-points in the jet chamber,
a momentum component perpendicular to the beam axis of
at least 0.15 GeV/$c$,
an impact parameter relative to the primary event vertex
of less than 0.5 cm in the $r-\phi$ plane and 20 cm along the
$z$ direction.
In addition, the d$E$/d$x$ measurement must have a probability
greater than 1\% for the pion hypothesis.
The \piz\  and $\eta$ selections are improved by using
\begin{eqnarray}
 \Ptpim & = & \frac{ \Ptpi }
     { \Ptpi  + ( 1 - \Ptpi  )
   \frac{\sqrt{2\pi}\sigma}{\Delta m}
        \exp(\frac{1}{2}(\frac{m-m_0}{\sigma})^2)  }
\end{eqnarray}
instead of \iPtpi\  as the selection variable,
using the same values of $m_0$, $\sigma$ and $\Delta m$ as in
eq.~\ref{eqn-ptpi}.
The $\gamma\gamma$, $\gamma\gamma_c$ and $\gamma_c\gamma_c$ samples 
are simply summed because their purities are approximately
the same for a given value of \Ptpim.
The resolution on the \piz\  and $\eta$ momenta and energies
are improved by constraining their invariant masses
to their nominal values~\cite{bib-pdg} using a kinematic fit.
The four-momenta of the charged and neutral particles are added,
and the invariant mass of the system is evaluated.

%%%%% Je me suis couche ici
\subsection{Determination of the particle yields}
\label{sect-yield}

As in the previous section, the determination of the meson
yields is repeated using different data samples and
analysis procedures and the results are averaged to obtain
the central values.
Similarly, the fits are systematically repeated using different values of
cuts on \Ptpim, different parameterizations of both the background and
the signal and using different Monte Carlo samples
for the determination of the efficiency and, in
some cases, the shape of the background.
For channels involving a \piz, the cut on \Ptpim\  
is varied among the values
0.1, 0.2, 0.3, 0.5 and 0.7. For channels involving an $\eta$ in
the final state, the cut values are 0.05, 0.1, 0.2, 0.3 and 0.4.
Since the largest deviations in the predicted efficiencies observed in the
previous two sections are those obtained when comparing the two JETSET samples,
only these two Monte Carlo samples are used in this section.
In the following, the variations in the parameterization of the signals
and the backgrounds are presented channel by channel.

\subsection{Analysis of the $\pi^0\pi^{\pm}$ invariant mass spectra}
\label{sect-fitrho}

\begin{figure}[tbp]
\epsfig{file=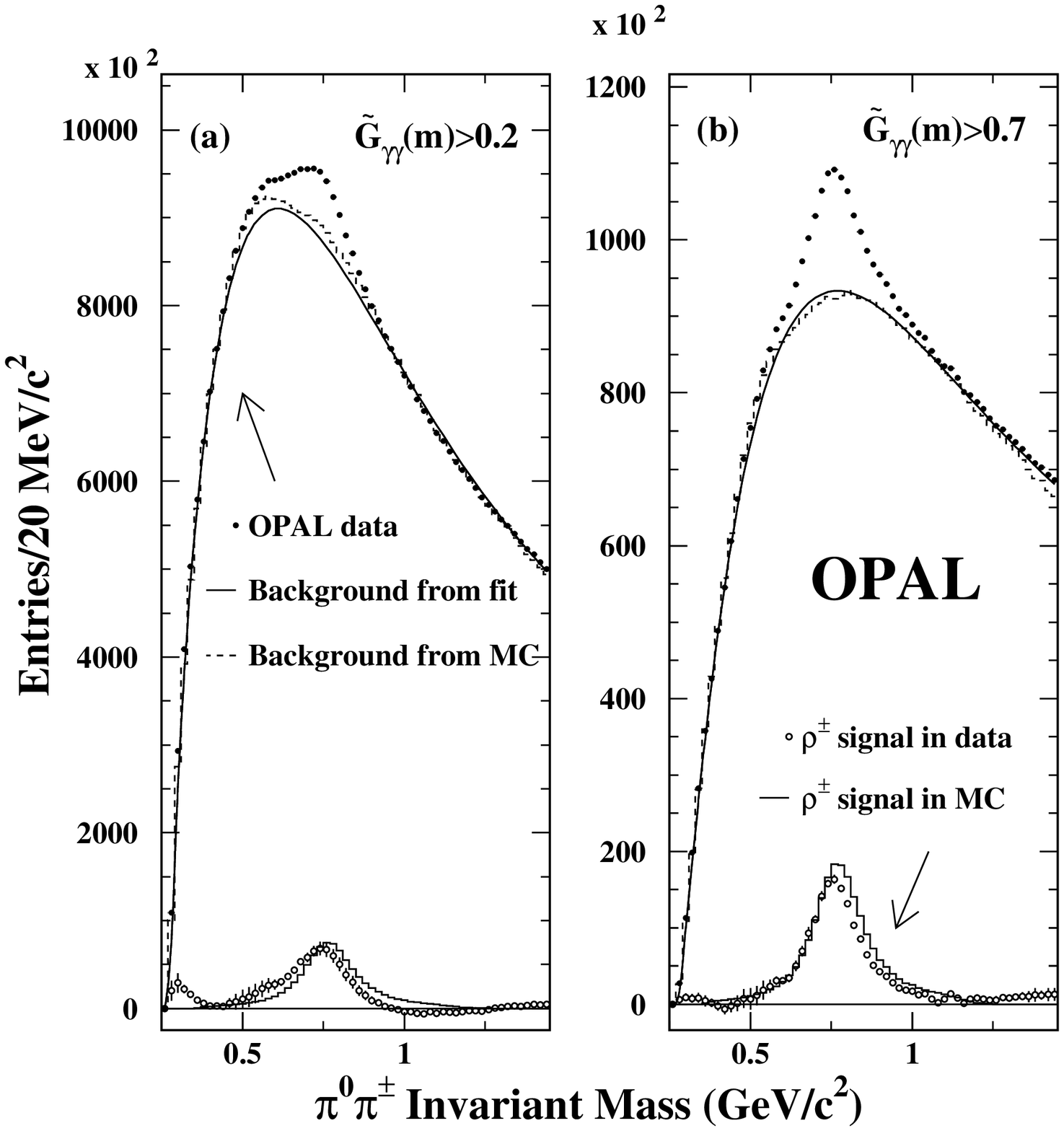
,height=17cm,bbllx=0cm,bblly=0cm,bburx=21cm,bbury=21cm}
\caption{ \label{fig-fitrho}
Fits to the invariant mass distribution of $\pi^0\pi^{\pm}$
combinations for two different cuts on \Ptpim.
The points represent the OPAL data and the full
curves the background obtained in the fit to the data.
The dashed histograms correspond to the background evaluated
using the Monte Carlo distributions.
In the lower part of the figures,
the signal obtained by subtracting the average of the
two background from the data
is shown as the open circles, and the signal
in the Monte Carlo, normalised to the same number of
events, is shown as the full histogram.}
\end{figure}

Fig.~\ref{fig-fitrho} shows the invariant mass spectra of
$\pi^0\pi^{\pm}$ combinations for the entire energy range
for two different values of the cut on \Ptpim.
The extraction of the \rpm\  yield from the $\pi^0\pi^{\pm}$
invariant mass spectra is complicated by the large width of
the resonance, by uncertainties regarding its exact shape
and by the reflection from $\omega\rightarrow\pi^0\pi^+\pi^-$ decays.
An additional complication is the presence of partially reconstructed
\rpm\  decays,
where only the charged pion and one of the photons come from the decay of
a \rpm\  meson while the other photon candidate has another origin.
It is therefore particularly useful to compare yields obtained
using different cuts on \Ptpim\
since this variation produces significant shifts in the position
of the maximum of the background shape (fig.~\ref{fig-fitrho})
and also changes the relative number of partially reconstructed \rpm\
decays.

Two methods are used to evaluate the background.
In the first, the background shape is taken from a simulation.
It is normalised to the number of counts outside the signal region.
This method is used to obtain background-subtracted invariant mass
spectra.
In the second method, the background is parameterized as:
\begin{eqnarray} \label{eqn-rhobkg}
   f(m) & = & p_1 (\Delta_m)^{p_2} \times \exp( p_3\Delta_m + p_4 \Delta_m^2 )
   \mbox{ , }
\end{eqnarray}
where $\Delta_m$ = $m-m_{\pi^0}-m_{\pi^{\pm}}$,
$m$ is the invariant mass of the $\pi^0\pi^{\pm}$ system
and the parameters $p_1$ to $p_4$ are determined in the fits to the data.
A Gaussian representing the reflection from \omegapipipi\  decays
is added to this shape, with a width fixed to the Monte Carlo prediction.
The amplitude and centroid of the Gaussian are free parameters,
in order to absorb possible imperfections in the modelling of the
background in this region close to the $\pi^0\pi^{\pm}$
threshold.
The simulation predicts that the reflections from \etapipipi\  and
$\mathrm{K}^*(892)^{\pm}\rightarrow\mathrm{K}^{\pm}\pi^0$ decays
are small; they are not included in the fit.
The background shapes obtained using the two methods
are shown in fig.~\ref{fig-fitrho}.
Also shown are the data before and after the subtraction of
the average of the two background shapes.
The \rpm\   resonance is clearly observed, albeit at a slightly lower
mass than in the Monte Carlo.

The shape of the \rpm\  resonance may be more complex than
a relativistic Breit-Wigner.
This is the case for the $\rho^0$ meson~\cite{bib-delrho,bib-opalrho},
where the observed deviations from this shape may
be due to residual Bose-Einstein correlations,
to interference between the amplitudes of the $\rho$ and the $\pi\pi$
background and to interference with the $\omega$~\cite{bib-lafferty}.
The first two effects should also affect the \rpm.
The most apparent sign of this distortion is a shift
towards low mass of the maximum of the resonance.
For this reason 
the position of the pole of the resonance is a free parameter
in the fits.
Following ref.~\cite{bib-lafferty}, the modification of the
$\rho$ shape is taken into account by multiplying the
relativistic Breit-Wigner by a factor
\begin{eqnarray}
I(m,C) & = &  1 \; + \; C \; \frac{m_0^2-m^2}{m\Gamma}
 \mbox{ , }
\end{eqnarray}
where the parameter $C$ is to be determined from the data.
The values of $C$ obtained in fits to the data
in different energy bins vary between 0.2 and 1.0, 
corresponding to shifts
of $-10$ to $-30$ MeV/$c^2$
in the position of the maximum of the resonance and
consistent with the observations of
ref.~\cite{bib-lafferty} for the $\rho^0$.

The \rpm\  yields are evaluated either from the results of
fits to the invariant mass spectra or by numerical integration
of background-subtracted spectra.
In the fits, the resonance is parameterized as a relativistic Breit-Wigner
convoluted with the experimental mass resolution and multiplied by
the factor $I(m,C)$.
The fits are repeated with the parameter $C$ being either
fixed to zero or left as a free parameter.
In addition, the experimental resolution is either fixed to the
Monte Carlo prediction or left as a free parameter.
These variations affect the number of mesons contained in
the tails of the resonance.
To address this problem in a consistent way,
the numbers of \rpm\  mesons are obtained by integrating the fitted
resonance shapes over the same range used for the integration of
the background-subtracted spectra, from 0.39 to 1.15 GeV/$c^2$.
All numbers in the data are then multiplied by 1.08 $\pm$ 0.03,
which is the
average correction needed to extrapolate the fitted shapes to
the range from the $\pi^0\pi^{\pm}$ threshold up to 1.5 GeV/$c^2$.
The integration range corresponds to the mass range to which 
the resonance is truncated in JETSET and therefore the yields
extracted from the simulations are not multiplied by this factor.
In addition, JETSET uses a non-relativistic Breit-Wigner shape,
without any of the correlations or interference effects just discussed
and for this reason the shape of the resonance is determined
separately for the data and the Monte Carlo samples.

Another important consideration in the fits to the invariant mass
spectra is the treatment of partially reconstructed \rpm\  mesons.
If a photon related to the \rpm\  decay contains most of the energy of the
\piz\   and is combined with any low-energy photon, the invariant mass
distribution of the system will form a broad peak under the signal.
The height of this ``bump'' is correlated with the \rpm\  yield,
but it is wider than the peak of fully reconstructed \rpm\  mesons.
The influence of partially reconstructed \rpm\   mesons is
taken into account in the analysis and in the systematic errors
in the following ways.
The bias in the extracted yields is evaluated by comparing the results
of fits to the invariant mass spectra of Monte Carlo samples where the\
partially reconstructed \rpm\   are included or not.
In addition,
the simulation predicts that the importance of partial reconstruction
decreases as the cut on \Ptpim\  (and therefore the quality of
the selected photons) increases.
The comparison of the rates measured with different cut values
is thus an additional test of how well this effect is simulated.

\subsection{Analysis of the $\eta\pi^{\pm}$ invariant mass spectra}

\begin{figure}[tbp]
\epsfig{file=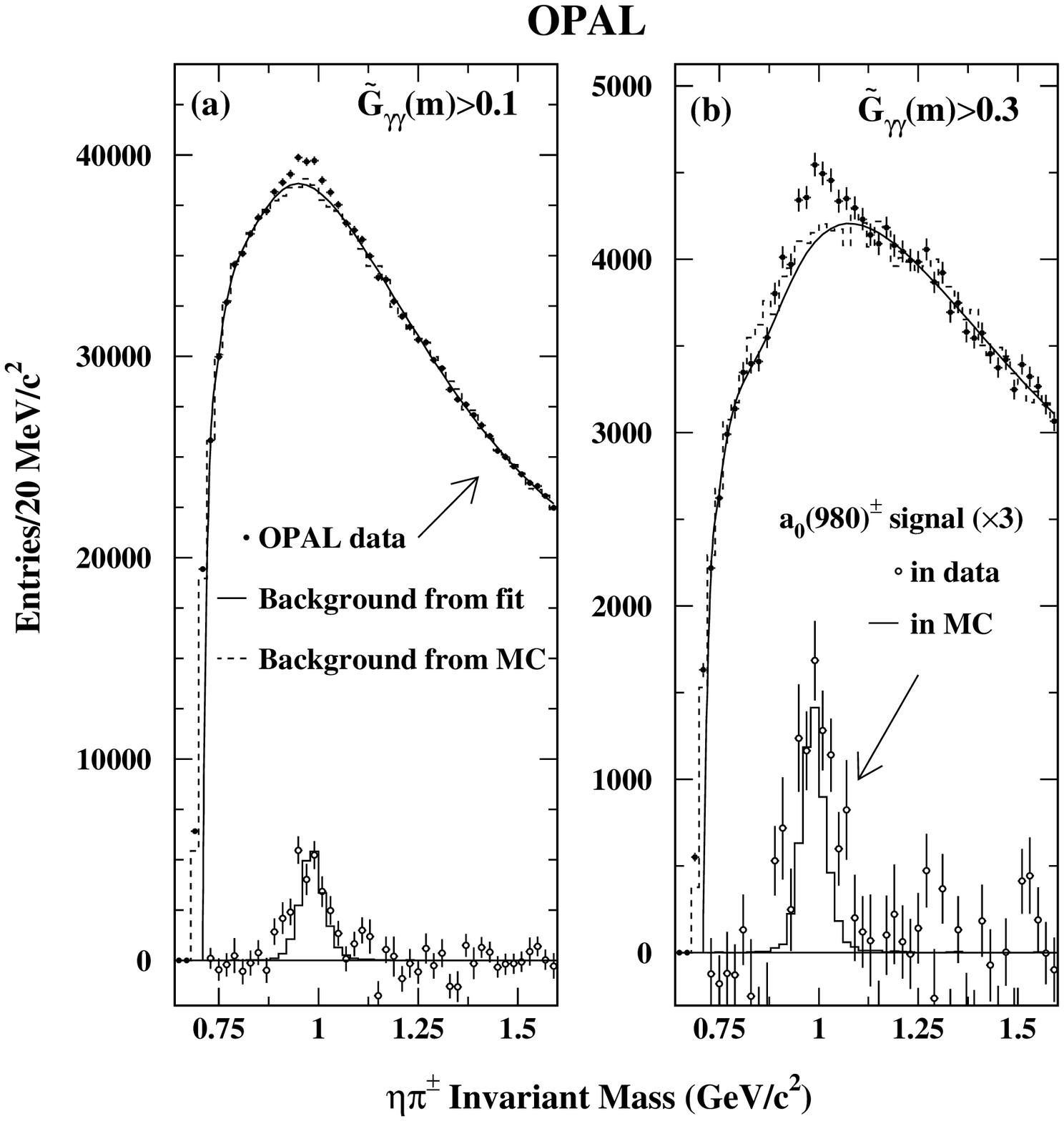
,height=17cm,bbllx=0cm,bblly=0cm,bburx=21cm,bbury=21cm}
\caption{ \label{fig-fitaz}
Fits to the invariant mass distribution of $\eta\pi^{\pm}$
combinations for two different cuts on \Ptpim.
The points represent the OPAL data and the full
curves the background obtained in the fit to the data.
The dashed histograms correspond to the background evaluated
using the Monte Carlo distributions.
In the lower part of the figures,
the signal obtained by subtracting the average of the two
background from the data
is shown as the open circles and is
multiplied by a factor 3 for clarity.
The full histogram is the Monte Carlo signal
normalised to the same number of events.}
\end{figure}

Fig.~\ref{fig-fitaz} shows the invariant mass spectra of
$\eta\pi^{\pm}$ combinations for the entire energy range
for two values of the cut on \Ptpim.
The analyses of the $\eta\pi^{\pm}$ and $\pi^0\pi^{\pm}$ 
invariant mass spectra are similar.
The same parameterization for the combinatorial background is used
(eq.~\ref{eqn-rhobkg})
and a Gaussian is added to represent the reflection from \etapetapipi\  decays.
However, because of the low statistics the description in the fit
of the \azpm\   resonance is not as detailed as for the \rpm.
The \azpm\  peak is parameterized by a simple Gaussian.
The fitted mean and width of the signal peak are 990 $\pm$ 12
and 51 $\pm$ 9 MeV/$c^2$, respectively,
in agreement with the nominal mass of 983.5~$\pm$~0.9 MeV/$c^2$~\cite{bib-pdg},
and consistent with the expected range of 50 to 90 MeV/$c^2$ for the width of
the resonance in the $\eta\pi^{\pm}$ channel~\cite{bib-azwid}.
To control possible biases due to the Gaussian assumption,
the fitted yields are used in the averaging
procedure together with the integrals of background-subtracted
spectra in the range from 880 to 1120 MeV/$c^2$,
where the background shape is either taken from the simulation
or from a fit of the analytical background shape to the data
outside the signal region.

In fig.~\ref{fig-fitaz}, the data are shown together with two
background distributions obtained either by the fit to the data
or by scaling the Monte Carlo prediction.
Also shown is the data after the average of the two background shapes
is subtracted.
A peak is observed, with a position and a width comparable
to the prediction of the Monte Carlo for the \azpm\
signal.
In the JETSET Monte Carlo the \azpm\
resonance is a non-relativistic Breit-Wigner with a pole at
$m_0$~=~983~MeV/$c^2$ and a width $\Gamma$~=~57~MeV/$c^2$, 
the distribution being truncated at $m_0$~$\pm$~50 MeV/$c^2$.
Due to the severe truncation and
taking into account the experimental mass resolution,
the simulated signal distribution shown in fig.~\ref{fig-fitaz}
resembles closely that of a Gaussian of width $\sigma$ $\sim$ 32 MeV/$c^2$.
According to the simulation, the effect of partially reconstructed
$\eta$ mesons is negligible.
This is because the $\eta$ selection imposes more stringent
requirements on the photon quality than does the \piz\   selection.

\subsection{Analysis of the $\pi^0\pi^+\pi^-$ invariant mass spectra}
\label{sect-fitom}

\begin{figure}[tbp]
\epsfig{file=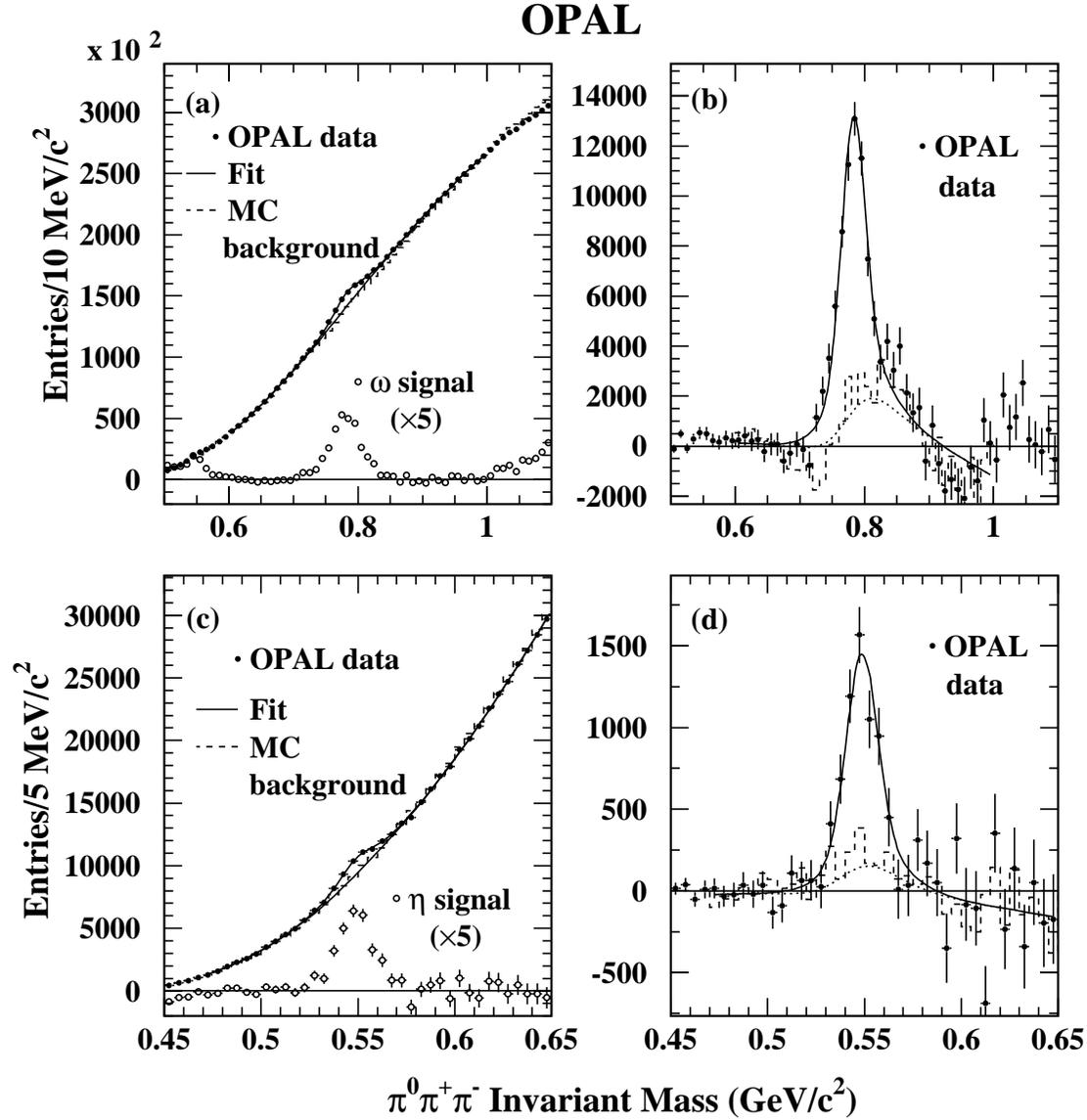
,height=17cm,bbllx=0cm,bblly=0cm,bburx=21cm,bbury=21cm}
\caption{ \label{fig-fitom}
Fits to the invariant mass distribution of $\pi^0\pi^+\pi^-$
combinations, for a cut of \Ptpim$>0.3$. 
The points represent the OPAL data and the full
curves the fit to the data.
The dashed histograms correspond to the background evaluated
using the Monte Carlo distributions.
Fig. (a) and (c): fits in the region
of the $\omega$ and the $\eta$, respectively.
The signal extracted from the fit to the data,
multiplied by a factor 5 for clarity,
is shown by the open circles.
Fig. (b) and (d): the points are the components proportional
to $\lambda$ (see section~\protect\ref{sect-fitom}) extracted
from the data
in the region of the $\omega$ and the $\eta$, respectively.
The full curves are the fits to the data.
The dashed histogram is the simulated background,
including partially reconstructed mesons which
appear as a bump close to the signal peak.
The dotted curves are the contributions from partial
reconstructions included in the fit.}
\end{figure}

Fig.~\ref{fig-fitom}a and c show the invariant mass spectra of \threepi
combinations, for the entire energy range, in the region
close to the $\eta$ and $\omega$ signals.
Compared to the $\pi^0\pi^{\pm}$ invariant mass spectra,
the $\pi^0\pi^+\pi^-$ analysis benefits from the narrow
widths of the $\eta$ and $\omega$ states.
However, it suffers from the reduced meson rates and the additional
combinatorial background.

In the fits to the invariant mass distributions,
the peaks from $\pi^0\pi^+\pi^-$ decays of the $\eta$ and
the $\omega$ are each well reproduced by two Gaussians sharing
the same centroid, their relative widths and areas
being determined from the simulation.
In the case of the $\omega$, this double-Gaussian also helps to
account for the tails of the Breit-Wigner distribution of the
resonance.
The fitted parameters of the peak are the area, its centroid
and its rms width.

The combinatorial backgrounds for the $\eta$ and the $\omega$ are
described by a second- and a third-order polynomial, respectively.
The shape of the fit component representing partially
reconstructed mesons differs from the one used for the \rpm.
The $\eta$ and $\omega$ momenta are shared amongst three
daughters instead of two as in the case of the \rpm.
Consequently,
the cases where all particles except one of the two photons come
from the same decay result in a bump with a narrower width,
lying systematically at masses above the peak of
fully-reconstructed decays. 
The adopted shape is an exponential tail at high masses
convoluted with a Gaussian
with a width equal to that of the signal peak.
The exponent of the tail and its area relative to the
signal peak are taken from the Monte Carlo prediction.

Two tests are performed to verify that the bump due to partially
reconstructed decays is properly described by the Monte Carlo.
The first is to compare the rates obtained with different
cuts on \Ptpim.
As the \piz\  probability increases,
the size of the bump relative to the signal decreases.
The second test uses the characteristic matrix elements of the
$\eta$ and $\omega$ decays to produce invariant mass spectra
with almost no background.
Within an experimental precision of a few percent~\cite{bib-dkmatrix},
the decay transition probabilities $\lambda_{\eta}$ and $\lambda_{\omega}$
are proportional to\footnote{As formulated here, $\lambda_{\eta}$ must
be multiplied by the $T^*_0$ phase-space dependence.}:
\begin{eqnarray}
\lambda_{\eta}   & \propto & 1 - \frac{T^*_0}{T^*_{0,max}}    \\
\lambda_{\omega} & \propto & |\vec{p}^{\,*}_{-}\times\vec{p}^{\,*}_{+}|^2
  \mbox{ , }
\end{eqnarray}
where $T^*_0$ is the kinetic energy of the \piz\  in the \threepi\
rest frame, $T^*_{0,max}$ is its maximum possible value and
$\vec{p}^{\,*}_{+}$ ($\vec{p}^{\,*}_{-}$) is the momentum of the positively
(negatively) charged pion in the \threepi\  rest frame.
$\lambda_{\eta}$ and $\lambda_{\omega}$ are normalised such that
they vary from 0 to 1.
Random combinations of three pions distributed according to
phase space result in a flat $\lambda$ distribution.
Therefore the ratio of the $\lambda$ distributions for the signal and
the combinatorial background should be proportional to $\lambda$.
Extracting from the invariant mass spectra the component
proportional to $\lambda$, one obtains distributions with
the combinatorial background subtracted, and where the shape of the
signal and of the bump of partially reconstructed mesons
can be evaluated with more precision.

The method to extract the component proportional to $\lambda$ 
in the data does not depend on its modelling in the Monte Carlo.
The $\lambda$ distribution of the candidates is sampled as a function
of their invariant mass, $m$.
The behaviour of phase space, $F(\lambda,m)$, including acceptance
effects, is taken from combinations with invariant masses above
and below the mass peak. In the signal region, it is interpolated using
a polynomial function.
In each invariant mass bin, the data are fitted with two components,
one proportional to the phase space behaviour, $F(\lambda,m)$, and 
the other proportional to $\lambda F(\lambda,m)$.
Figs.~\ref{fig-fitom}b and d show the components proportional
to  $\lambda F(\lambda,m)$ extracted from the data in
figs.~\ref{fig-fitom}a and c, respectively.
The distribution for the $\omega$ signal is slightly asymmetric due
to a small excess at high mass which is explained in the Monte Carlo
as the bump of partially reconstructed mesons (dashed line in
fig.~\ref{fig-fitom}b).
Tests performed on the simulations show that
the components proportional to $\lambda F(\lambda,m)$
are excellent approximations of the signal distributions.
In both data and Monte Carlo, the meson yields obtained
from fits to the total invariant mass distribution or to the
component proportional to $\lambda$ agree within 3\%.

In JETSET, the matrix element of the decay \etapipipi\
is not simulated.
To reproduce the data, the signal events in the Monte Carlo
are weighted by a factor\footnote{The normalisation is chosen such
that the integral of the signal, $\int_0^1 F(\lambda,m)d\lambda$, is 1.}
of 2$\lambda_{\eta}$.
Due to the strong correlation of $\lambda_{\eta}$ with the \piz\
energy, this correction changes the total efficiency for
the detection of $\eta\rightarrow\pi^0\pi^+\pi^-$ decays
by as much as 20\%.

\subsection{Analysis of the $\eta\pi^+\pi^-$ invariant mass spectra}
\label{sect-etapfit}
      
\begin{figure}[tbp]
\epsfig{file=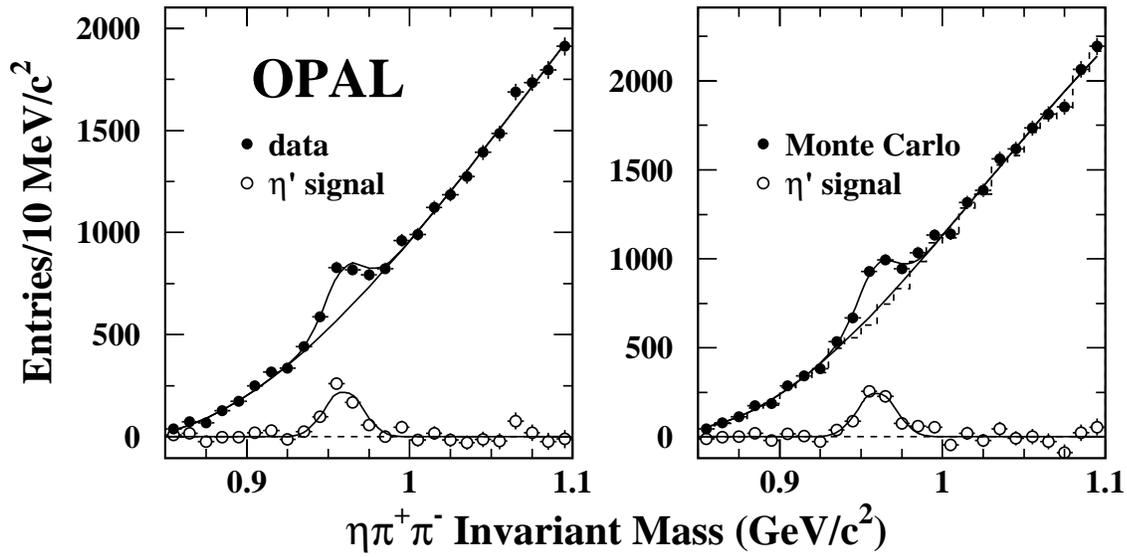
,height=10cm,bbllx=0cm,bblly=9cm,bburx=21cm,bbury=21cm}
\caption{ \label{fig-fitetap}
Fit to the invariant mass distribution of $\eta\pi^+\pi^-$
combinations in the data (left) and in the Monte Carlo
(right).
A cut on \Ptpim$>0.1$ has been applied.
The curves represent the fit and 
the Monte Carlo background is shown as a dashed histogram.
In the lower part of the figures, the open symbols
represent the signal extracted from the fit.}
\end{figure}

Fits to the $\eta\pi^+\pi^-$ invariant mass spectra for
the entire energy range are shown in fig.~\ref{fig-fitetap},
for the data and the Monte Carlo.
Given the small statistics,
a Gaussian is found to describe adequately the $\eta'$ signal.
As for the \azpm, the contribution from partially reconstructed
$\eta$ decays does not need to be parameterized in the fit.
In contrast to the three-body decays of the $\eta$ and $\omega$,
the Dalitz plot for the decay \etapetapipi\  is closer to phase
space~\cite{bib-epmatrix} and
the matrix element of the decay cannot
be used to obtain background-free distributions.

%%%%%%%%%%%%%%%%%%%%%%%%%%%%%%%%%%%%%%%%%%%%%%%%%%%%%%%%%%%%%%%%%%%%%%%%%%%%

\subsection{ Determination of the meson yields}
\label{sect-xsec}

The \rpm, \azpm, $\eta$, $\omega$ and $\eta'$ yields and their systematic
errors are determined using the same averaging method as in
sections~\ref{sect-grate} and~\ref{sect-rategg}.
The yields are corrected for the known branching ratios of the different
decay modes~\cite{bib-pdg},
except for \azetapi, where a branching ratio of 90~$\pm$~10~\%
is assumed~\cite{bib-pdg,bib-azbnl}.

The numbers of mesons per hadronic \Zzero\  decay
in the energy ranges covered by the present measurement
are given in table~\ref{tab-chgerr} together with the values
of each systematic uncertainty.
These are:
\begin{itemize}
\item{The statistical error on the Monte Carlo samples used to
      calculate the efficiency.}
\item{The variations observed using different Monte Carlo samples,
      obtained from the averaging procedure.}
\item{The error associated with \Ptpim,
      obtained from the averaging procedure.}
\item{The variations observed when using the different background
      parameterizations,
      obtained from the averaging procedure.
      In the case of $\pi^0\pi^+\pi^-$ decays, this includes the
      variations observed when the matrix elements $\lambda_{\eta}$ and
      $\lambda_{\omega}$ are used.}
\item{The variations observed when the mass resolution is
      either fixed to the Monte Carlo prediction or left as a free
      parameter.
      In the case of the \rpm, this also includes the variations
      observed when $C$ is either fixed to zero or left as 
      a free parameter.}
\item{For the \rpm, the error associated with the extrapolation of
      the resonance beyond the mass range
      described in section~\ref{sect-fitrho}.}
\item{The simulation uncertainties not covered by the variations
      of the cut on \Ptpim. The contribution of the \piz\  and
      $\eta$ are the simulation errors quoted in tables~\ref{tab-pizerr}
      and~\ref{tab-etaerr}.
      These are added in quadrature and correspond to a 2\% uncertainty
      on the charged pion selection.}
\item{The uncertainty due to partially reconstructed mesons, evaluated
      with the Monte Carlo by comparing the results of fits to mass
      distributions with and without their contribution.
      The yield difference is taken as a systematic error.}
\item{In the case of the decays $\eta\rightarrow\pi^0\pi^+\pi^-$,
      $\eta'\rightarrow\eta\pi^+\pi^-$ and
      \azpm$\rightarrow\eta\pi^{\pm}$, the uncertainty on the
      branching ratio is greater than 1\%, and is therefore included.}
\end{itemize}

The differential cross sections as a function of \xE\ and \lnxp\  are
presented, interpreted and discussed together with those for the photons
and the other mesons in section~\ref{sect-results}. 

%%%%%%%%%%%%%%%%%%%%%%%%%%%%%%%%%%%%%%%%%%%%%%%%%%%%%%%%%%
%%%%%%%%%%%%%%%%%%%%%%%%%%%%%%%%%%%%%%%%%%%%%%%%%%%%%%%%%%

\section{Combination of channels}
\label{sect-combi}

The calorimeter and conversion data are compared.
The ratio of the total photon rates obtained using identified photon
conversions and the calorimeter data is 1.010 $\pm$ 0.002 (stat.),
in agreement well within the size of the systematic errors on the two
samples (\gamerrpct\% and \gamcerrpct\%; see table~\ref{tab-gamerr}).
Comparison of the \piz\  rates measured in the energy range
common to the three channels $\gamma\gamma$, $\gamma\gamma_c$ and
$\gamma_c\gamma_c$ also yields ratios consistent with one.
The ratios of the rates $\gamma\gamma_c$/$\gamma\gamma$ and 
$\gamma_c\gamma_c$/$\gamma\gamma$ are 0.97 $\pm$ 0.02 (stat.) and
0.96 $\pm$ 0.05 (stat.), respectively,
in perfect agreement given the estimated \piggerrpct\%,
\piggcerrpct\% and \pigcgcerrpct\%
errors on the rates for the $\gamma\gamma$,
$\gamma\gamma_c$ and $\gamma_c\gamma_c$ channels, respectively
(table~\ref{tab-pizerr}).
The agreement is also excellent for the $\gamma\gamma_c$
and $\gamma\gamma$ channels for the $\eta$, despite a
\etaggcerrpct\% uncertainty on the $\gamma\gamma_c$ channel
(table~\ref{tab-etaerr}).
 
Given the positive result of these consistency tests,
a weighted average of the differential cross-sections measured
using the different channels is performed.
The weights are taken as the inverse of the square of the total
errors. The systematic errors on the combined results are obtained
by assuming that the individual systematic errors are completely
correlated.
In the few cases where the individual measurements and their errors
% are not consistent with the averaged result,
are more than one standard deviation away from the averaged result,
the error on the average is scaled using the method of
ref.~\cite{bib-pdg}.

The $\eta$ differential cross section measurements
based on the $\gamma\gamma(\gamma\gamma_c)$ and
$\pi^0\pi^+\pi^-$ channels have comparable total errors
(\etaggterrpct\% and \etappperrpct\%, respectively;
see tables~\ref{tab-etaerr} and ~\ref{tab-chgerr})
and the systematic uncertainties are largely uncorrelated.
The two measurements are combined assuming that the errors associated
to the track and cluster simulation and to the  variations of \Ptpim\  
and of the Monte Carlo are entirely correlated and that all other sources of
uncertainty are uncorrelated.
In the energy range where both types of data are available,
the ratio of the \threepi\  to the   $\gamma\gamma(\gamma\gamma_c)$
results is constant and equal to 1.14 $\pm$ 0.07 (stat.) $\pm$ 0.13 (syst.).
With the improved knowledge of the absolute normalisation provided
by the combination of the two data sets,
the error on the total $\eta$ rate is 10.9\%.

\section{Results}
\label{sect-results}

In this section, the experimental meson rates are presented
and compared to the predictions of the JETSET and HERWIG
Monte Carlo models.
To simplify the comparison, the predictions
of the default versions of JETSET 7.4 and HERWIG 5.9 are
used here, except for the \azpm\   which is not produced
in the default version of JETSET 7.4 and for which the
prediction of the JETSET version of ref.~\cite{bib-tunenew}
is used.
Two aspects of the model predictions are investigated:
the shape of the momentum spectra and the integrated rates.

\subsection{Differential cross-sections}
\label{sect-difxsec}

\begin{figure}[tbp]
\epsfig{file=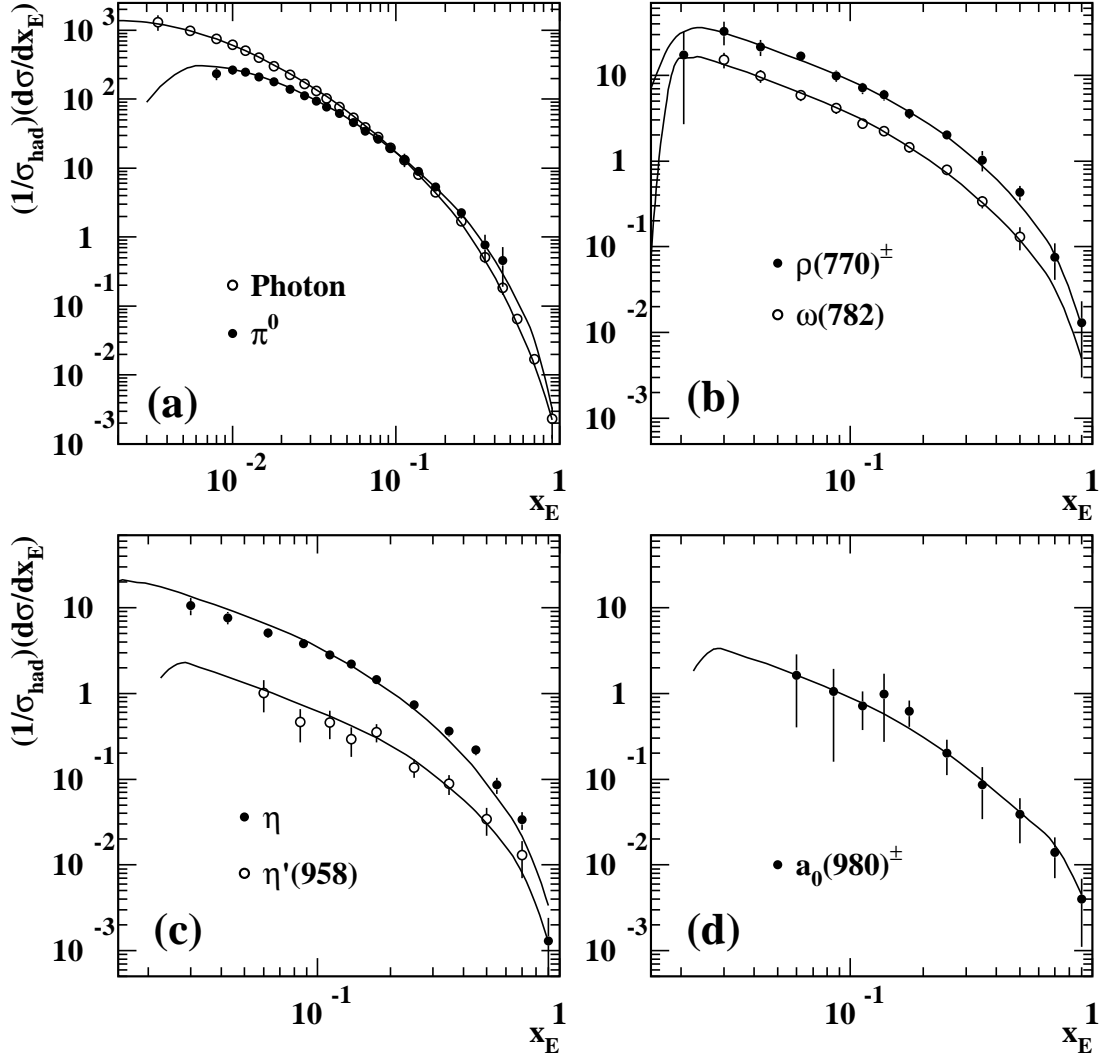
,height=17cm,bbllx=0cm,bblly=0cm,bburx=21cm,bbury=21cm}
\caption{ \label{fig-fragxe} 
Differential cross-section as a function of the
scaled energy \xE\  for the production in hadronic \Zzero\
decays of (a) photons and \piz\  mesons,
(b) \rpm\  and $\omega$ mesons,
(c) $\eta$ and $\eta'$ mesons and
(d) \azpm\   mesons.
The points are the data and the curves are the predictions
of default JETSET 7.4, with a normalisation fitted to the data.
The error bars represent the quadratic sum of the statistical
and systematic errors.}
\end{figure}

The differential cross-sections as a function of \xE\  and \lnxp\
of the photon and the 
\piz, $\eta$, \rpm, $\omega$, $\eta'$ and \azpm\  mesons 
are obtained by dividing the yields by the corresponding bin widths.
For the \rpm\  and \azpm\  resonances, the relationship between
the meson energy and its momentum varies from event to event.
The yields are thus evaluated separately in bins of \xE\  and
in bins of \lnxp.
The differential cross-sections are listed
in tables~\ref{tab-fragg} to~\ref{tab-fragaz}
together with the statistical and systematic errors.
In fig.~\ref{fig-fragxe}, the data are compared to
the JETSET 7.4 predictions normalised to the measured rate.
In the simulation, the difference between
the slopes of the photon and \piz\  distributions
(fig.~\ref{fig-fragxe}a) is strongly constrained by the
fact that 92\% of the photons come from \piz\  decays.
The measured shapes are consistent with the photon and
\piz\  predictions.
The \rpm\  and $\omega$ mesons, both vector particles
with similar masses, have similar slopes
(fig.~\ref{fig-fragxe}b) 
that are also well reproduced by the Monte Carlo.
However the predicted $\eta$ spectrum is too soft
compared to the measurement (fig.~\ref{fig-fragxe}c),
while the $\eta'$ prediction is consistent with the
data within errors.
The slope of the \azpm\  distribution (fig.~\ref{fig-fragxe}d)
is well reproduced by JETSET with the parameters of
ref.~\cite{bib-tunenew}.

\begin{figure}[tbp]
\epsfig{file=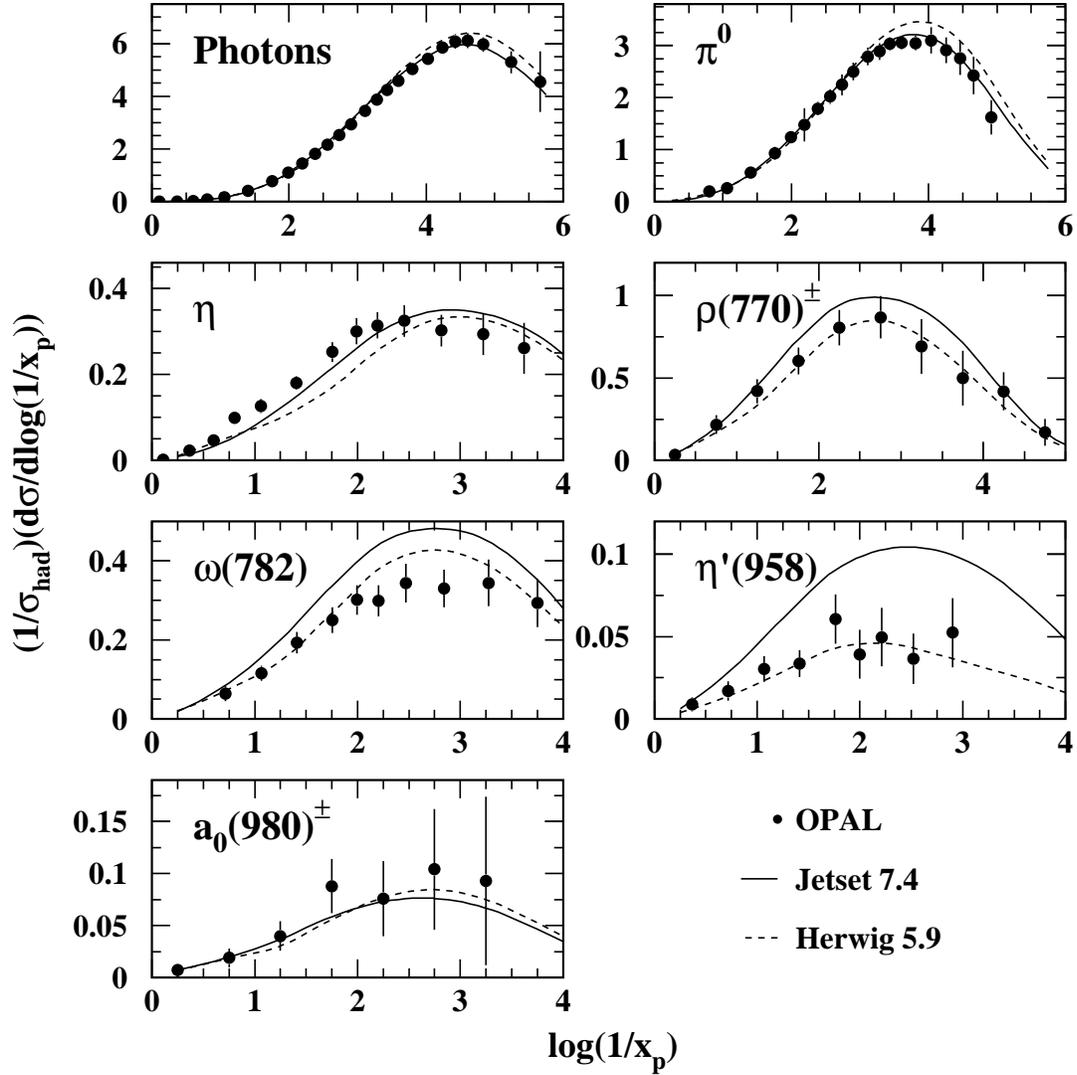
,height=17cm,bbllx=0cm,bblly=0cm,bburx=21cm,bbury=21cm}
\caption{ \label{fig-fragln} OPAL measurement
of the inclusive production of photons, \piz\, $\eta$,
\rpm\, $\omega$, $\eta'$ and \azpm\   mesons in
hadronic \Zzero\   decays, as a function of \lnxp.
The full and dashed lines are the absolute predictions of
JETSET 7.4 and HERWIG 5.9, respectively, without
any additional normalisation.
The error bars represent the quadratic sum of the statistical
and systematic errors.}
\end{figure}

  To emphasise the low-momentum portions of
the spectra which represent the largest fraction of the
inclusive rates, the differential cross-sections
are presented in fig.~\ref{fig-fragln} as a function of \lnxp.
In this figure, the full and dashed curves are the absolute
predictions of default JETSET 7.4 and HERWIG 5.9, respectively,
except for the \azpm\  prediction of JETSET, taken from
ref.~\cite{bib-tunenew}.
  JETSET 7.4 reproduces the photon and \piz\  data slightly
better than HERWIG 5.9.
The predictions for the \azpm\  are quite similar, but
HERWIG 5.9 provides a better description of the \rpm, $\omega$
and $\eta'$ data.
However, the $\eta$ momentum spectra predicted by both models
are too soft.

\subsection{Maxima of the $\log(1/x_p)$ distributions}

The location of the maximum of the \lnxp\   distribution
is expected to be correlated with the mass of the particle~\cite{bib-thln}.
In addition, its value provides a quantitative measurement of
the hardness of the momentum spectrum.
The value is  extracted by fitting a Gaussian to the data close to
the maximum of the distribution.
These values are listed in table~\ref{tab-lnxpmax},
and shown in fig.~\ref{fig-tab}a
where they are compared to other measurements at LEP
and to the predictions of JETSET and HERWIG.
The errors quoted in the table are the sum of the fit
errors and of the uncertainties due to variations of the range
of the Gaussian fits.

As can be seen in fig.~\ref{fig-tab}a,
both JETSET and HERWIG reproduce the data within errors,
except for the $\eta$, for which the maximum is at
a higher value in both JETSET and HERWIG.
This confirms that the $\eta$ spectrum in the two models is too soft
as was already noted in section~\ref{sect-difxsec}.
The OPAL results agree with those of
DELPHI~\cite{bib-delpiz} and
L3~\cite{bib-lpiz,bib-leta,bib-letap}.

\subsection{Rates extrapolated to $0<x_E<1$}

The extrapolation of the rates to the full range of \xE\
is done using the fractions of the rate lying outside the
measured range predicted by JETSET 7.4, HERWIG 5.9 and a Gaussian
fit to the $\log(1/x_p)$ distributions.
The applied corrections correspond to the average of the lowest
and highest values and the maximum
deviation is taken as the systematic error on the procedure.
The data used to evaluate the extrapolation factors and
the associated errors are listed in table~\ref{tab-extrap}.
The results of the Gaussian fits are considered because
of their good description of the data
within errors and because in at least one case (the $\eta$)
it appears that the JETSET and HERWIG shapes may not be
appropriate.
Other shapes that describe the data equally well are also
considered.
For the \piz\  and the $\eta$ they are,
respectively, the $\pi^{\pm}$ and K$^0_S$ distributions measured
by OPAL~\cite{bib-opalpkpi,bib-opalkz}.
The extrapolation factors derived from these experimental
shapes are all within the range covered by the predictions
of JETSET, HERWIG and the Gaussian fit.

The particle multiplicities per hadronic \Zzero\  decay
extrapolated to the full energy range are:
\begin{eqnarray}
\langle n_{\gamma} \rangle              & = & \gamresl \nonumber \mbox{ , } \\
\langle n_{\pi^0} \rangle               & = & \pizresl \nonumber \mbox{ , } \\
\langle n_{\eta} \rangle                & = & \etaresl \nonumber \mbox{ , } \\
\langle n_{\rho^{\pm}} \rangle          & = & \rhoresl \nonumber \mbox{ , } \\
\langle n_{\omega} \rangle              & = & \omeresl \nonumber \mbox{ , } \\
\langle n_{\eta'} \rangle               & = & \etpresl \nonumber \mbox{ , } \\
\langle n_{{\mathrm a}_0^{\pm}} \rangle & = & \azeresl \nonumber \mbox{ , }
\end{eqnarray}
where the first errors are statistical, the second systematic
and the third are from the extrapolation procedure.

\subsection{Discussion of the rates and comparison with models}

\begin{figure}[tbp]
\epsfig{file=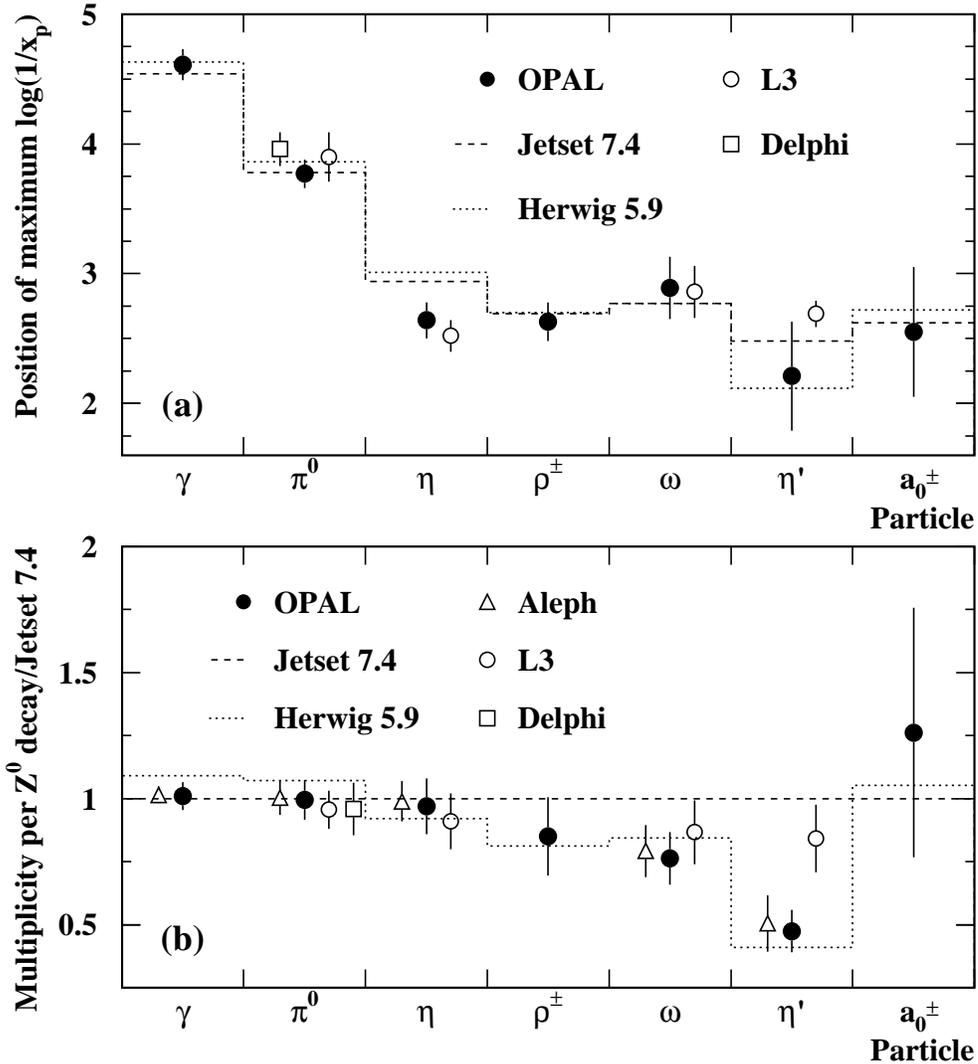
,height=17cm,bbllx=0cm,bblly=0cm,bburx=21cm,bbury=21cm}
\caption{ \label{fig-tab} 
a) Position of the maximum of the \lnxp\   distributions.
The black points are the OPAL measurements,
and the full and dashed lines are the predictions of
JETSET 7.4 and HERWIG 5.9, respectively.
The measurements of DELPHI~\protect\cite{bib-delpiz} and
L3~\protect\cite{bib-lpiz,bib-leta,bib-letap}
are shown as white squares
and circles, respectively.
The maxima are obtained by a Gaussian fit to the data
close to the maximum, except for L3, which makes
model-dependent assumptions concerning the shape
of the distribution.
b) Particle multiplicities (extrapolated to $0<x_E<1$),
divided by the prediction of JETSET 7.4.
The full circles represent the OPAL measurements, and the open
circles, triangles and squares those of other LEP
experiments~\protect\cite{bib-delpiz,bib-alcomp,bib-lpiz,bib-leta,bib-letap}.
The dotted line represents the prediction of HERWIG 5.9.}
\end{figure}

In table~\ref{tab-others}, the measured rates are compared to those
from other LEP experiments and to the predictions of JETSET 7.4
and HERWIG 5.9.
The ratio of the measured rates to the JETSET 7.4 predictions
are shown in fig.~\ref{fig-tab}b together with the results from other LEP
experiments~\cite{bib-alcomp,bib-delpiz,bib-lpiz,bib-leta,bib-letap}.
In this figure, the results obtained by ALEPH in limited energy ranges
for the photon, $\eta$ and $\eta'$ are divided by
the JETSET 7.4 predictions in the corresponding range.
All the measurements are consistent with each other,
except perhaps for the $\eta'$ where the measured rate agrees with the
ALEPH result~\cite{bib-alcomp}, but is more than two standard
deviations away from that of L3~\cite{bib-letap}.

According to JETSET 7.4, 97.0\% of all photons come from
\piz, $\eta$, $\omega$ and $\eta'$ decays.
The prediction of HERWIG 5.9 is 96.0\%.
In comparison, the sum of the measured \piz, $\eta$, $\omega$ and $\eta'$
rates multiplied by the known photon multiplicities
in their decays~\cite{bib-pdg} accounts for (95 $\pm$ 5)\%
of the measured number of photons\footnote{
In the data, the \piz\  and $\eta$ decays alone account for approximately
91\% and 4\% of all observed photons, respectively, in good agreement
with the JETSET 7.4 predictions of 91.9\% and 4.0\%.}.
The error on the ratio is calculated assuming that the errors
on the photon, \piz\  and $\eta$ rates are fully correlated except
for those associated with the background subtraction and the fit to the
invariant mass spectra\footnote{These are the errors labelled as
``Background subtraction'', ``Gaussian peaks'' and
``Background normalisation range'' in tables 1 to 3.
The contributions of the $\omega$ and the $\eta'$ to the total
error are negligible.}.
This good agreement provides further evidence that the models offer a reasonable
description of the inclusive production of photons in \Zzero\  decays,
an assumption on which the measurement of the photon rates relies.

The production rates of the \piz\  and \rpm\  can be compared
with those of their isospin parters.
Using the measurements for the $\pi^{\pm}$~\cite{bib-opalpkpi}
and the $\rho^0$~\cite{bib-alcomp,bib-delrho}, the following ratios
of rates are obtained:
\begin{eqnarray}
2 \langle n_{\pi^0}  \rangle / \langle n_{\pi^{\pm}}  \rangle & = &
1.12 \pm 0.01 \pm 0.08 \nonumber \mbox{ , } \\
2 \langle n_{\rho^0} \rangle / \langle n_{\rho^{\pm}} \rangle & = &
1.08 \pm 0.04 \pm 0.20 \nonumber \mbox{ , }
\end{eqnarray}
where the first errors are statistical and the second systematic.
These ratios are consistent with the predictions of JETSET 7.4
(1.132 and 1.064, respectively) and HERWIG 5.9 (1.147 and 1.032).
In these models, most of the deviation from unity comes from
the decays $\eta\rightarrow\pi^0\pi^0\pi^0$ and
$\eta'\rightarrow\rho^0\gamma$.

As seen in fig.~\ref{fig-tab}b,
the measured rates are consistent with the predictions of JETSET 7.4
and HERWIG 5.9, except for the $\omega$ and the $\eta'$ for which
the rates are more than two standard deviations smaller than the
JETSET 7.4 prediction.
The failure of JETSET to reproduce the $\eta'$ rate is well
understood, as it assumes a similar strange quark content for
the $\eta$ and $\eta'$ and neglects the effect of their
difference in mass on their relative production rate.
For this reason, the suppression of the $\eta'$
relative to the $\eta$ is a free parameter in JETSET 7.4.
The present data suggest that the current suppression factor
of 0.4 should be further reduced.
In contrast to the $\eta'$, no single parameter can
modify the $\omega$ rate in JETSET independently of
all other mesons.
In that model an increase of the $\omega$ rate
is necessarily accompanied by an equivalent increase
of the $\rho^0$, $\rho^+$ and $\rho^-$ rates
because, with ideal mixing, these mesons are the corresponding
isospin $I=0$ and $I=1$ states.
Indeed, the measured \rpm\  and $\omega$ rates (table~\ref{tab-others})
are consistent with $I=0$/$I=1$ symmetry, albeit with a large error.
However that symmetry can be broken by
cascade decays of heavier mesons such as the $L=1$ states.
The experimental information on the production of these
states is limited and the present \azpm\  data is interesting
in this respect.

Table~\ref{tab-others} shows that the a$_0^+$ and  a$_0^-$ are
produced at rates comparable to the
$\eta'$, a meson of equal spin and similar mass.
However, the strangeness content of the \azpm\   and
$\eta'$ are not expected to be the same.
A more relevant comparison is with the f$_0$(980) meson,
which is the isospin $I$=0 partner of the \azpm\
according to the quark model of mesons.
The ratio of the rates of the \azpm\   and the
f$_0$(980)~\cite{bib-opalfz} is 1.9 $\pm$ 0.8,
compatible with the expected value of 2.

In the HERWIG cluster fragmentation model~\cite{bib-herwig},
the relative production rates of light-flavour mesons are
mostly determined by their masses, which affects the phase space
available for the cluster decay.
This simple ansatz appears to be able to reproduce the measured \azpm\
rate.
The \azpm\  is not present in the default version of JETSET.
The inclusion of $L=1$ meson production in hadronisation requires
the tuning of additional parameters.
The predictions shown in fig.~\ref{fig-tab} correspond
to the choice of parameters of ref.~\cite{bib-tunenew} optimised,
in part, to reproduce the available data on D and B mesons.
The agreement for the \azpm\   may be accidental since the
parameters of ref.~\cite{bib-tunenew} also predict a
substantial b$_1(1232)^{\pm,0}$ rate of 0.92 meson per
\Zzero\   decays.
The present $\omega$ and \rpm\   data do not support this
prediction. 
With the b$_1(1232)^{\pm,0}$ decaying exclusively to $\omega\pi$,
the parameters of ref.~\cite{bib-tunenew} result in an $\omega$
rate which exceeds that of its $I=1$ partner, the $\rho$,
with $\langle n_{\omega} \rangle-\langle n_{\rho^{\pm}} \rangle/2$ = 0.56.
The measured difference is $-0.17$~$\pm$~0.26, in better agreement
with the prediction of $-0.06$ of default JETSET 7.4
with no $L=1$ mesons, and with the prediction of
0.00 of HERWIG 5.9, with a b$_1(1232)^{\pm,0}$ rate of 0.32.
Thus, despite the lack of direct measurements for several $L=1$ meson
states, it appears possible to constrain the JETSET model
by using the available data on all other mesons.

%%%%%%%%%%%%%%%%%%%%%%%%%%%%%%%%%%%%%%%%%%%%%%%%%%%%%%%%%%%%%%%%%%%%%%%%%%%%

\section{Conclusion}
\label{sect-sum}

The inclusive particle multiplicity per hadronic \Zzero\  decay 
and the differential cross-section have been measured for photons
and for \piz, $\eta$, \rpm, $\omega$, $\eta'$ and \azpm\   mesons.
The \azpm\ is observed for the first time in high-energy
\epem\   collisions.
It is produced at a rate comparable to that of mesons with a
similar mass, such as the $\eta'$ and the $f_0(980)$.
The inclusive \rpm\  production is measured for the
first time in hadronic \Zzero\  decays.
The models JETSET 7.4 and HERWIG 5.9 with their default
parameters reproduce the shape of the measured differential
cross-sections, with the exception of that of the $\eta$ meson
which is too soft in both models.
The absolute rates in HERWIG 5.9 are in good agreement.
In JETSET 7.4, the production rates of the $\omega$ and
$\eta$ are overestimated by  20\% and 50\%, respectively.
The present \azpm\  data is a valuable input for
the determination of the parameters required for
the inclusion of $L=1$ mesons in JETSET.
These parameters are further constrained by
the data on other mesons like the $\omega$ and the
\rpm.
 
%%%%%%%%%%%%%%%%%%%%%%%%%%%%%%%%%%%%%%%%%%%%%%%%%%%%%%%%%%%%%%%%%%%%%%%%%%%%
\newpage
\noindent {\Large \bf Acknowledgements}
\par
\noindent We particularly wish to thank the SL Division for the efficient operation
of the LEP accelerator at all energies
 and for their continuing close cooperation with
our experimental group.  We thank our colleagues from CEA, DAPNIA/SPP,
CE-Saclay for their efforts over the years on the time-of-flight and trigger
systems which we continue to use.  In addition to the support staff at our own
institutions we are pleased to acknowledge the  \\
Department of Energy, USA, \\
National Science Foundation, USA, \\
Particle Physics and Astronomy Research Council, UK, \\
Natural Sciences and Engineering Research Council, Canada, \\
Israel Science Foundation, administered by the Israel
Academy of Science and Humanities, \\
Minerva Gesellschaft, \\
Benoziyo Center for High Energy Physics,\\
Japanese Ministry of Education, Science and Culture (the
Monbusho) and a grant under the Monbusho International
Science Research Program,\\
German Israeli Bi-national Science Foundation (GIF), \\
Bundesministerium f\"ur Bildung, Wissenschaft,
Forschung und Technologie, Germany, \\
National Research Council of Canada, \\
Research Corporation, USA,\\
Hungarian Foundation for Scientific Research, OTKA T-016660, 
T023793 and OTKA F-023259.\\
%%%%%%%%%%%%%%%%%%%%%%%%%%%%%%%%%%%%%%%%%%%%%%%%%%%%%%%%%%%%%%%%%%%%%%%%%%%%

%========================================================================
 
\newpage
 
%========================================================================

\begin{table}[htbp]
\begin{center}
\begin{tabular}{|l|c|c||c|}
\hline
\hline
 Photon sample      & $\gamma$ & $\gamma_c$ & $\gamma+\gamma_c$ \\
\hline
 \xE\  range   & 0.003 - 1.000  & 0.003 - 1.000 & 0.003 - 1.000  \\
\hline
 Integrated rate      & 16.79          & 16.96         &   16.84        \\
\hline
\hline
 Errors (\%)                  & \multicolumn{3}{c|}{\mbox{ }} \\
\hline
 Statistics (data)            &  0.1  &  0.2 &  0.1 \\
 Statistics (Monte Carlo)     &  0.1  &  0.1 &  0.1 \\
 Difference between Monte Carlos      &  3.4  &  3.0 &  3.3 \\
 $P_{\gamma}(S)$ variations   &  0.8  &   -  &  0.6 \\
 Background subtraction       &  0.8  &  0.5 &  0.7 \\
 Nuclear interactions         &  2.0  &  0.9 &  1.7 \\
 Energy scale                 &  1.5  &  1.3 &  1.4 \\
 Track and cal. simulation    &  1.8  &  7.6 &  3.1 \\
\hline
 Total error (\%)             &  \gamerrpct\
                              &  \gamcerrpct\
                              &  5.1 \\
\hline
\hline
\end{tabular}
\end{center}
\caption{\label{tab-gamerr}
Number of photons per hadronic \Zzero\   decay
in the \xE\   range covered by the measurement
together with its statistical
and systematic uncertainties (in \%).
The three columns are the results obtained
with the calorimetric sample ($\gamma)$,
the conversion sample ($\gamma_c$)
and the combined sample ($\gamma+\gamma_c$).}
\end{table}

%========================================================================

\begin{table}[htbp]
\begin{center}
\begin{tabular}{|l|c|c|c||c|}
\hline
\hline
 \piz\   sample   & $\gamma\gamma$ & $\gamma\gamma_c$ & 
                 $\gamma_c\gamma_c$ & 
      $\gamma\gamma+\gamma\gamma_c+\gamma_c\gamma_c$ \\
\hline
 \xE\   Range   & 0.007 - 0.500  & 0.007 - 0.500 & 0.009 - 0.300 & 0.007-0.500 \\
\hline
 Integrated rate &  8.37       &     8.12      &    7.51       &     8.29    \\
\hline
 Rate ($0.009<x_E<0.3$)
              &     7.80       &     7.35      &    7.51       &     7.65    \\
\hline
\hline
 Errors (\%)                  & \multicolumn{4}{c|}{\mbox{ }} \\
\hline
 Statistics (data)            &  0.6  &  2.1 &  4.9 &  0.6 \\
 Statistics (Monte Carlo)     &  0.4  &  1.0 &  2.8 &  0.4 \\
 Difference between Monte Carlos      &  4.7  &  4.0 &  5.9 &  4.1 \\
 \Ptpi\   variations          &  2.0  &  2.7 &  9.3 &  2.2 \\
 Background subtraction       &  3.7  &  7.2 & 11.5 &  3.8 \\
 Gaussian peaks               &  0.8  &  1.1 &  3.0 &  0.8 \\
 Nuclear interactions         &  1.6  &  0.7 &  0.1 &  1.3 \\
 Energy scale (1\%)           &  1.8  &  1.9 &  1.5 &  1.6 \\
 Background normalisation range &  4.3  &  2.3 &  5.0 &  3.1 \\
 Track and cal. simulation    &  2.2  &  7.6 & 15.2 &  2.5 \\
\hline
Total error (\%)              &  \piggerrpct\ 
                              &  \piggcerrpct\
                              &  \pigcgcerrpct
                              &  7.6 \\
\hline
\hline
\end{tabular}
\end{center}
\caption{\label{tab-pizerr}
Statistical and systematic uncertainties (in \%) on the
number of \piz\   mesons per hadronic \Zzero\   decay
measured using the individual \pigg, \piggc\  and
\pigcgc\  samples, and the combined sample.}
\end{table}

%========================================================================

\begin{table}[htbp]
\begin{center}
\begin{tabular}{|l|c|c||c|}
\hline
\hline
$\eta$ sample   & $\gamma\gamma$ & $\gamma\gamma_c$ & $\gamma\gamma+\gamma\gamma_c$ \\
\hline
\hline 
       \xE\   Range      & 0.025 - 1.000  & 0.040 - 0.300 & 0.025 - 1.000 \\
\hline
Integrated rate        &     0.746     &      0.569    &     0.745     \\
\hline
Rate ($0.04<x_E<0.30$) &     0.558     &      0.569    &     0.558     \\
\hline 
\hline 
  Errors (\%)          & \multicolumn{3}{c|}{\mbox{ }} \\
\hline
Statistics (data)            &  2.6  &  6.5 &  2.6 \\
Statistics (Monte Carlo)     &  1.7  &  4.5 &  1.7 \\
Difference between Monte Carlos      &  3.9  &  9.7 &  3.8 \\
\Ptpi\   variations          & 10.5  & 26.0 & 10.5 \\
Background subtraction       &  7.4  & 16.7 &  7.3 \\
Gaussian peaks               &  0.2  &  1.1 &  0.2 \\
Energy scale (1\%)           &  0.8  &  0.4 &  0.8 \\
Background normalisation range &  1.9  &  3.9 &  1.9 \\
Track and cal. simulation    &  1.1  &  6.9 &  1.1 \\
\hline
Total Error (\%)             & \etaggerrpct\  
                             & \etaggcerrpct\  
                             & \etaggterrpct\  \\
\hline
\hline
\end{tabular}
\end{center}
\caption{\label{tab-etaerr}
Statistical and systematic uncertainties (in \%) on the
number of $\eta$ mesons per hadronic \Zzero\   decay
measured using the individual \etagg\  and
\etaggc\  samples, and the combined sample.}
\end{table}

%========================================================================

\begin{table}[htbp]
\begin{center}
\begin{tabular}{|l|c|c|c|c|c|}
\hline
 Meson       & $\eta$    & \rpm\   & $\omega$ & $\eta'$ & \azpm\   \\
 Decay  mode & \threepi\   & $\pi^0\pi^{\pm}$ &
               \threepi\   & $\eta\pi^+\pi^-$ & $\eta\pi^{\pm}$ \\
\hline
 \xE\   range     & 0.025-0.4 &         & 0.025-0.6 & 0.05-0.8 &        \\
\hline
 \lnxp\   range   &           & 0.0-5.0 &           &          & 0.0-3.5 \\
\hline
 Rate             & 0.898     & 2.36    & 0.883     & 0.103    & 0.214   \\
\hline
\hline
 Errors (\%)                  & \multicolumn{5}{c|}{\mbox{ }} \\
\hline
 Statistics (data)            &  6.5 &  2.3 &  3.6 & 10.1 & 16.8 \\
 Statistics (Monte Carlo)     &  1.7 &  1.7 &  0.5 &  1.6 &  2.5 \\
 Difference between Monte Carlos      &  3.7 &  -   &  5.9 &  4.9 &  -   \\
 \Ptpim\   variations         &  2.4 & 10.9 &  3.9 &  7.2 & 13.6 \\
 Background subtraction       &  4.5 &  7.4 &  3.0 &  5.0 & 30.4 \\
 Mass resolution              &  4.0 &  7.0 &  8.1 &  6.8 &  -   \\
 Breit-Wigner extrapolation   &   -  &  3.1 &   -  &   -  &  -   \\
 Track and cal. simulation    &  4.0 &  4.0 &  4.0 &  3.0 &  3.0 \\   
 Partial reconstruction       &  6.6 &  9.0 &  5.8 &  -   &  6.6 \\
 Branching Ratio              &  2.5 &   -  &   -  &  3.4 & 10.3 \\
\hline
 Total Error (\%)             & \etappperrpct\ & 18.4 & 13.7 & 16.5 & 39.4 \\
\hline
\hline
\end{tabular}
\end{center}
\caption{\label{tab-chgerr}
Statistical and systematic uncertainties (in \%) on the
number of $\eta$, \rpm, $\omega$, $\eta'$ and \azpm\   mesons
per hadronic \Zzero\   decay measured with the channels combining 
a \piz\   or an $\eta$ meson with charged pions.
For the \rpm\   and \azpm\   resonances, the error associated
with the difference between the Monte Carlos is included in
the error associated with the background subtraction.}
\end{table}

%========================================================================

\clearpage

%========================================================================

\begin{table}[htbp]
\begin{center}
\begin{tabular}{|c|c||c|c|}
\hline
   \xE\       & $\frac{1}{\sigma_{had}}\frac{d\sigma}{dx_E}$           &
 ln$(1/x_p)$  & $\frac{1}{\sigma_{had}} \frac{d\sigma}{d{\mathrm ln}(1/x_p)}$    \\
\hline
%              gamerr.com
\hline
 0.003 - 0.004 &   1309 $\pm$     13 $\pm$    330 &  5.81 -  5.52 &   4.55 $\pm$   0.05 $\pm$   1.15 \\
 0.004 - 0.007 &    986 $\pm$      2 $\pm$     78 &  5.52 -  4.96 &   5.29 $\pm$   0.01 $\pm$   0.42 \\
 0.007 - 0.009 &    749 $\pm$      1 $\pm$     34 &  4.96 -  4.71 &   5.96 $\pm$   0.01 $\pm$   0.27 \\
 0.009 - 0.011 &    613 $\pm$      1 $\pm$     28 &  4.71 -  4.51 &   6.11 $\pm$   0.01 $\pm$   0.28 \\
 0.011 - 0.013 &    508 $\pm$      1 $\pm$     23 &  4.51 -  4.34 &   6.08 $\pm$   0.02 $\pm$   0.27 \\
 0.013 - 0.016 &    404 $\pm$      1 $\pm$     16 &  4.34 -  4.14 &   5.84 $\pm$   0.01 $\pm$   0.23 \\
 0.016 - 0.020 &    303 $\pm$      1 $\pm$     11 &  4.14 -  3.91 &   5.43 $\pm$   0.01 $\pm$   0.20 \\
 0.020 - 0.025 &    225 $\pm$      1 $\pm$      9 &  3.91 -  3.69 &   5.03 $\pm$   0.01 $\pm$   0.19 \\
 0.025 - 0.030 &    167 $\pm$      1 $\pm$      7 &  3.69 -  3.51 &   4.58 $\pm$   0.01 $\pm$   0.18 \\
 0.030 - 0.035 &    131 $\pm$      1 $\pm$      5 &  3.51 -  3.35 &   4.24 $\pm$   0.01 $\pm$   0.15 \\
 0.035 - 0.040 &    103 $\pm$      1 $\pm$      4 &  3.35 -  3.22 &   3.87 $\pm$   0.01 $\pm$   0.15 \\
 0.040 - 0.050 &   76.9 $\pm$    0.2 $\pm$    2.8 &  3.22 -  3.00 &   3.45 $\pm$   0.01 $\pm$   0.13 \\
 0.050 - 0.060 &   53.4 $\pm$    0.2 $\pm$    2.0 &  3.00 -  2.81 &   2.93 $\pm$   0.01 $\pm$   0.11 \\
 0.060 - 0.070 &   39.2 $\pm$    0.2 $\pm$    1.4 &  2.81 -  2.66 &   2.54 $\pm$   0.01 $\pm$   0.09 \\
 0.070 - 0.085 &   28.1 $\pm$    0.1 $\pm$    1.0 &  2.66 -  2.47 &   2.17 $\pm$   0.01 $\pm$   0.08 \\
 0.085 - 0.100 &   19.6 $\pm$    0.1 $\pm$    0.8 &  2.47 -  2.30 &   1.81 $\pm$   0.01 $\pm$   0.08 \\
 0.100 - 0.125 &   13.0 $\pm$    0.1 $\pm$    0.5 &  2.30 -  2.08 &   1.45 $\pm$   0.01 $\pm$   0.06 \\
 0.125 - 0.150 &   8.04 $\pm$   0.05 $\pm$   0.35 &  2.08 -  1.90 &  1.103 $\pm$  0.006 $\pm$  0.048 \\
 0.150 - 0.200 &   4.50 $\pm$   0.03 $\pm$   0.26 &  1.90 -  1.61 &  0.783 $\pm$  0.005 $\pm$  0.045 \\
 0.200 - 0.300 &   1.71 $\pm$   0.02 $\pm$   0.19 &  1.61 -  1.20 &  0.422 $\pm$  0.004 $\pm$  0.047 \\
 0.300 - 0.400 &  0.507 $\pm$  0.010 $\pm$  0.087 &  1.20 -  0.92 &  0.176 $\pm$  0.003 $\pm$  0.030 \\
 0.400 - 0.500 &  0.184 $\pm$  0.005 $\pm$  0.038 &  0.92 -  0.69 &  0.082 $\pm$  0.002 $\pm$  0.017 \\
 0.500 - 0.600 &  0.065 $\pm$  0.002 $\pm$  0.011 &  0.69 -  0.51 &  0.036 $\pm$  0.001 $\pm$  0.006 \\
 0.600 - 0.800 &  0.017 $\pm$  0.001 $\pm$  0.002 &  0.51 -  0.22 &  0.012 $\pm$  0.000 $\pm$  0.002 \\
 0.800 - 1.000 & 0.0023 $\pm$ 0.0003 $\pm$ 0.0010 &  0.22 -  0.00 & 0.0020 $\pm$ 0.0003 $\pm$ 0.0009 \\
\hline
\hline
\end{tabular}
\end{center}
\caption{\label{tab-fragg} Photon fragmentation function
obtained by combining the calorimeter and conversion results.
The quoted errors are statistical and systematic, respectively. }
\end{table}

%========================================================================

\begin{table}[htbp]
\begin{center}
\begin{tabular}{|c|c||c|c|}
\hline
   \xE\      & $\frac{1}{\sigma_{had}}\frac{d\sigma}{dx_E}$           &
 ln$(1/x_p)$ & $ \frac{1}{\sigma_{had}} \frac{d\sigma}{d{\mathrm ln}(1/x_p)}$   \\
\hline
%              pizerr.com
\hline
 0.007 - 0.009 &    254 $\pm$     18 $\pm$     48 &  5.06 -  4.77 &   1.74 $\pm$   0.12 $\pm$   0.33 \\
 0.009 - 0.011 &    266 $\pm$     12 $\pm$     38 &  4.77 -  4.55 &   2.42 $\pm$   0.11 $\pm$   0.34 \\
 0.011 - 0.013 &    248 $\pm$      6 $\pm$     28 &  4.55 -  4.37 &   2.78 $\pm$   0.06 $\pm$   0.31 \\
 0.013 - 0.016 &    211 $\pm$      3 $\pm$     18 &  4.37 -  4.15 &   2.92 $\pm$   0.05 $\pm$   0.25 \\
 0.016 - 0.020 &    178 $\pm$      2 $\pm$     14 &  4.15 -  3.92 &   3.11 $\pm$   0.03 $\pm$   0.25 \\
 0.020 - 0.025 &    139 $\pm$      2 $\pm$      6 &  3.92 -  3.70 &   3.05 $\pm$   0.03 $\pm$   0.13 \\
 0.025 - 0.030 &    113 $\pm$      1 $\pm$      5 &  3.70 -  3.51 &   3.06 $\pm$   0.03 $\pm$   0.13 \\
 0.030 - 0.035 &   94.1 $\pm$    0.9 $\pm$    4.0 &  3.51 -  3.36 &   3.03 $\pm$   0.03 $\pm$   0.13 \\
 0.035 - 0.040 &   77.7 $\pm$    0.8 $\pm$    4.3 &  3.36 -  3.22 &   2.89 $\pm$   0.03 $\pm$   0.16 \\
 0.040 - 0.050 &   62.5 $\pm$    0.4 $\pm$    3.9 &  3.22 -  3.00 &   2.79 $\pm$   0.02 $\pm$   0.17 \\
 0.050 - 0.060 &   45.7 $\pm$    0.4 $\pm$    3.0 &  3.00 -  2.81 &   2.50 $\pm$   0.02 $\pm$   0.16 \\
 0.060 - 0.070 &   34.7 $\pm$    0.3 $\pm$    3.0 &  2.81 -  2.66 &   2.25 $\pm$   0.02 $\pm$   0.20 \\
 0.070 - 0.085 &   26.2 $\pm$    0.2 $\pm$    1.8 &  2.66 -  2.47 &   2.02 $\pm$   0.02 $\pm$   0.14 \\
 0.085 - 0.100 &   19.4 $\pm$    0.2 $\pm$    1.4 &  2.47 -  2.30 &   1.79 $\pm$   0.02 $\pm$   0.13 \\
 0.100 - 0.125 &   13.2 $\pm$    0.1 $\pm$    2.9 &  2.30 -  2.08 &   1.48 $\pm$   0.02 $\pm$   0.32 \\
 0.125 - 0.150 &   9.05 $\pm$   0.13 $\pm$   0.76 &  2.08 -  1.90 &  1.240 $\pm$  0.017 $\pm$  0.105 \\
 0.150 - 0.200 &   5.36 $\pm$   0.10 $\pm$   0.69 &  1.90 -  1.61 &  0.931 $\pm$  0.017 $\pm$  0.120 \\
 0.200 - 0.300 &   2.26 $\pm$   0.13 $\pm$   0.38 &  1.61 -  1.20 &  0.558 $\pm$  0.031 $\pm$  0.094 \\
 0.300 - 0.400 &  0.764 $\pm$  0.085 $\pm$  0.309 &  1.20 -  0.92 &  0.266 $\pm$  0.030 $\pm$  0.107 \\
 0.400 - 0.500 &  0.455 $\pm$  0.095 $\pm$  0.244 &  0.92 -  0.69 &  0.204 $\pm$  0.043 $\pm$  0.110 \\
\hline
\hline
\end{tabular}
\end{center}
\caption{\label{tab-fragpi} \piz\   fragmentation function
obtained by combining the \pigg, \piggc\   and \pigcgc\   data.
The quoted errors are statistical and systematic, respectively. }
\end{table}

%========================================================================

\begin{table}[htbp]
\begin{center}
\begin{tabular}{|c|c||c|c|}
\hline
   \xE\       & $\frac{1}{\sigma_{had}}\frac{d\sigma}{dx_p}$           &
 ln$(1/x_p)$  & $\frac{1}{\sigma_{had}} \frac{d\sigma}{d{\mathrm ln}(1/x_p)}$    \\
\hline
\hline
 0.025 - 0.035 &   10.6 $\pm$    1.5 $\pm$    2.4 &  3.82 -  3.42 &  0.261 $\pm$  0.038 $\pm$  0.046 \\
 0.035 - 0.050 &   7.63 $\pm$   0.78 $\pm$   1.27 &  3.42 -  3.03 &  0.294 $\pm$  0.030 $\pm$  0.038 \\
 0.050 - 0.075 &   5.10 $\pm$   0.38 $\pm$   0.61 &  3.03 -  2.60 &  0.302 $\pm$  0.023 $\pm$  0.028 \\
 0.075 - 0.100 &   3.81 $\pm$   0.21 $\pm$   0.44 &  2.60 -  2.31 &  0.324 $\pm$  0.018 $\pm$  0.032 \\
 0.100 - 0.125 &   2.83 $\pm$   0.12 $\pm$   0.28 &  2.31 -  2.08 &  0.314 $\pm$  0.014 $\pm$  0.028 \\
 0.125 - 0.150 &   2.21 $\pm$   0.10 $\pm$   0.22 &  2.08 -  1.90 &  0.301 $\pm$  0.014 $\pm$  0.027 \\
 0.150 - 0.200 &   1.46 $\pm$   0.05 $\pm$   0.13 &  1.90 -  1.61 &  0.252 $\pm$  0.009 $\pm$  0.021 \\
 0.200 - 0.300 &  0.733 $\pm$  0.026 $\pm$  0.062 &  1.61 -  1.20 &  0.180 $\pm$  0.006 $\pm$  0.014 \\
 0.300 - 0.400 &  0.364 $\pm$  0.022 $\pm$  0.047 &  1.20 -  0.92 &  0.126 $\pm$  0.008 $\pm$  0.014 \\
 0.400 - 0.500 &  0.220 $\pm$  0.019 $\pm$  0.031 &  0.92 -  0.69 &  0.099 $\pm$  0.008 $\pm$  0.011 \\
 0.500 - 0.600 &  0.086 $\pm$  0.010 $\pm$  0.019 &  0.69 -  0.51 &  0.047 $\pm$  0.006 $\pm$  0.009 \\
 0.600 - 0.800 &  0.033 $\pm$  0.004 $\pm$  0.008 &  0.51 -  0.22 &  0.023 $\pm$  0.003 $\pm$  0.005 \\
 0.800 - 1.000 & 0.0013 $\pm$ 0.0004 $\pm$ 0.0011 &  0.22 -  0.00 & 0.0012 $\pm$ 0.0004 $\pm$ 0.0009 \\
\hline
\hline
\end{tabular}
\end{center}
\caption{\label{tab-etacomb} $\eta$ fragmentation function obtained
by combining the $\gamma\gamma$, $\gamma\gamma_c$ and $\pi^0\pi^+\pi^-$
data. The quoted errors are statistical and systematic, respectively. }
\end{table}

%========================================================================

\begin{table}[htbp]
\begin{center}
\begin{tabular}{|c|c||c|c|}
\hline
   \xE\       & $\frac{1}{\sigma_{had}}\frac{d\sigma}{dx_p}$           &
 ln$(1/x_p)$  & $\frac{1}{\sigma_{had}} \frac{d\sigma}{d{\mathrm ln}(1/x_p)}$    \\
\hline
%              rhoerr.com
\hline
 0.016 - 0.025 &   17.3 $\pm$    8.1 $\pm$   12.2 & 5.0 - 4.5 &  0.171 $\pm$  0.008 $\pm$  0.081 \\
 0.025 - 0.035 &   32.3 $\pm$    2.5 $\pm$    9.7 & 4.5 - 4.0 &  0.419 $\pm$  0.035 $\pm$  0.111 \\
 0.035 - 0.050 &   21.3 $\pm$    0.7 $\pm$    4.5 & 4.0 - 3.5 &  0.500 $\pm$  0.092 $\pm$  0.138 \\
 0.050 - 0.075 &   16.7 $\pm$    0.4 $\pm$    1.8 & 3.5 - 3.0 &  0.692 $\pm$  0.028 $\pm$  0.165 \\
 0.075 - 0.100 &   9.89 $\pm$   0.40 $\pm$   1.46 & 3.0 - 2.5 &  0.868 $\pm$  0.021 $\pm$  0.126 \\
 0.100 - 0.125 &   7.11 $\pm$   0.25 $\pm$   1.04 & 2.5 - 2.0 &  0.805 $\pm$  0.022 $\pm$  0.104 \\
 0.125 - 0.150 &   5.90 $\pm$   0.25 $\pm$   0.78 & 2.0 - 1.5 &  0.603 $\pm$  0.017 $\pm$  0.078 \\
 0.150 - 0.200 &   3.60 $\pm$   0.12 $\pm$   0.48 & 1.5 - 1.0 &  0.419 $\pm$  0.014 $\pm$  0.073 \\
 0.200 - 0.300 &   2.02 $\pm$   0.07 $\pm$   0.21 & 1.0 - 0.5 &  0.217 $\pm$  0.010 $\pm$  0.055 \\
 0.300 - 0.400 &   1.03 $\pm$   0.04 $\pm$   0.27 & 0.5 - 0.0 &  0.034 $\pm$  0.004 $\pm$  0.019 \\
 0.400 - 0.600 &  0.430 $\pm$  0.023 $\pm$  0.081 &           &                                  \\
 0.600 - 0.800 &  0.075 $\pm$  0.013 $\pm$  0.032 &           &                                  \\
 0.800 - 1.000 &  0.013 $\pm$  0.003 $\pm$  0.009 &           &                                  \\
\hline
\hline
\end{tabular}
\end{center}
\caption{\label{tab-fragrho} \rpm\   fragmentation function.
The quoted errors are statistical and systematic, respectively.
Because of the width of the \rpm, the relation between \xE\    and
\xp\   varies with mass.
Therefore the analysis is performed first with bins of \xE\  
(first two columns) and then repeated with bins of \lnxp.}
\end{table}

%========================================================================

\begin{table}[htbp]
\begin{center}
\begin{tabular}{|c|c||c|c|}
\hline
   \xE\       & $\frac{1}{\sigma_{had}}\frac{d\sigma}{dx_p}$           &
 ln$(1/x_p)$  & $\frac{1}{\sigma_{had}} \frac{d\sigma}{d{\mathrm ln}(1/x_p)}$    \\
\hline
%              omeerr.com
\hline
 0.025 - 0.035 &   15.2 $\pm$    2.4 $\pm$    2.1 &  4.01 -  3.49 &  0.293 $\pm$  0.046 $\pm$  0.040 \\
 0.035 - 0.050 &   9.88 $\pm$   0.84 $\pm$   1.48 &  3.49 -  3.06 &  0.344 $\pm$  0.029 $\pm$  0.051 \\
 0.050 - 0.075 &   5.82 $\pm$   0.35 $\pm$   0.75 &  3.06 -  2.62 &  0.330 $\pm$  0.020 $\pm$  0.043 \\
 0.075 - 0.100 &   4.12 $\pm$   0.25 $\pm$   0.54 &  2.62 -  2.32 &  0.344 $\pm$  0.021 $\pm$  0.045 \\
 0.100 - 0.125 &   2.74 $\pm$   0.16 $\pm$   0.32 &  2.32 -  2.09 &  0.299 $\pm$  0.018 $\pm$  0.035 \\
 0.125 - 0.150 &   2.23 $\pm$   0.14 $\pm$   0.24 &  2.09 -  1.90 &  0.301 $\pm$  0.018 $\pm$  0.032 \\
 0.150 - 0.200 &   1.45 $\pm$   0.09 $\pm$   0.17 &  1.90 -  1.61 &  0.250 $\pm$  0.016 $\pm$  0.029 \\
 0.200 - 0.300 &  0.789 $\pm$  0.049 $\pm$  0.099 &  1.61 -  1.21 &  0.193 $\pm$  0.012 $\pm$  0.024 \\
 0.300 - 0.400 &  0.335 $\pm$  0.037 $\pm$  0.042 &  1.21 -  0.92 &  0.116 $\pm$  0.013 $\pm$  0.014 \\
 0.400 - 0.600 &  0.130 $\pm$  0.027 $\pm$  0.028 &  0.92 -  0.51 &  0.064 $\pm$  0.013 $\pm$  0.014 \\
\hline
\hline
\end{tabular}
\end{center}
\caption{\label{tab-fragom} $\omega$ fragmentation function.
The quoted errors are statistical and systematic, respectively.}
\end{table}

%========================================================================

\begin{table}[htbp]
\begin{center}
\begin{tabular}{|c|c||c|c|}
\hline
   \xE\       & $\frac{1}{\sigma_{had}}\frac{d\sigma}{dx_p}$           &
 ln$(1/x_p)$  & $\frac{1}{\sigma_{had}} \frac{d\sigma}{d{\mathrm ln}(1/x_p)}$    \\
\hline
%              etperr.com
\hline
 0.050 - 0.070 &   1.01 $\pm$   0.38 $\pm$   0.14 &  3.09 -  2.71 &  0.052 $\pm$  0.020 $\pm$  0.007 \\
 0.070 - 0.100 &  0.462 $\pm$  0.180 $\pm$  0.073 &  2.71 -  2.33 &  0.036 $\pm$  0.014 $\pm$  0.006 \\
 0.100 - 0.125 &  0.460 $\pm$  0.144 $\pm$  0.082 &  2.33 -  2.09 &  0.050 $\pm$  0.016 $\pm$  0.009 \\
 0.125 - 0.150 &  0.293 $\pm$  0.099 $\pm$  0.049 &  2.09 -  1.91 &  0.039 $\pm$  0.013 $\pm$  0.007 \\
 0.150 - 0.200 &  0.354 $\pm$  0.068 $\pm$  0.054 &  1.91 -  1.61 &  0.061 $\pm$  0.012 $\pm$  0.009 \\
 0.200 - 0.300 &  0.137 $\pm$  0.028 $\pm$  0.017 &  1.61 -  1.21 &  0.034 $\pm$  0.007 $\pm$  0.004 \\
 0.300 - 0.400 &  0.088 $\pm$  0.020 $\pm$  0.011 &  1.21 -  0.92 &  0.030 $\pm$  0.007 $\pm$  0.004 \\
 0.400 - 0.600 &  0.034 $\pm$  0.010 $\pm$  0.006 &  0.92 -  0.51 &  0.017 $\pm$  0.005 $\pm$  0.003 \\
 0.600 - 0.800 &  0.013 $\pm$  0.006 $\pm$  0.003 &  0.51 -  0.22 &  0.009 $\pm$  0.004 $\pm$  0.002 \\
\hline
\hline
\end{tabular}
\end{center}
\caption{\label{tab-fragetap} $\eta'$ fragmentation function.
The quoted errors are statistical and systematic, respectively.}
\end{table}

%========================================================================

\begin{table}[htbp]
\begin{center}
\begin{tabular}{|c|c||c|c|}
\hline
   \xE\       & $\frac{1}{\sigma_{had}}\frac{d\sigma}{dx_p}$           &
 ln$(1/x_p)$  & $\frac{1}{\sigma_{had}} \frac{d\sigma}{d{\mathrm ln}(1/x_p)}$    \\
\hline
%              azeerr.com
\hline
 0.050 - 0.070 &   1.65 $\pm$   1.03 $\pm$   0.75 & 3.50 -  3.00 &  0.093 $\pm$  0.063 $\pm$  0.050 \\
 0.070 - 0.100 &   1.05 $\pm$   0.49 $\pm$   0.73 & 3.00 -  2.50 &  0.104 $\pm$  0.041 $\pm$  0.041 \\
 0.100 - 0.125 &  0.747 $\pm$  0.215 $\pm$  0.214 & 2.50 -  2.00 &  0.076 $\pm$  0.019 $\pm$  0.030 \\
 0.125 - 0.150 &  0.985 $\pm$  0.238 $\pm$  0.560 & 2.00 -  1.50 &  0.088 $\pm$  0.013 $\pm$  0.023 \\
 0.150 - 0.200 &  0.623 $\pm$  0.107 $\pm$  0.171 & 1.50 -  1.00 &  0.040 $\pm$  0.009 $\pm$  0.012 \\
 0.200 - 0.300 &  0.207 $\pm$  0.046 $\pm$  0.069 & 1.00 -  0.50 &  0.019 $\pm$  0.006 $\pm$  0.007 \\
 0.300 - 0.400 &  0.093 $\pm$  0.027 $\pm$  0.040 & 0.50 -  0.00 & 0.0071 $\pm$ 0.0025 $\pm$ 0.0022 \\
 0.400 - 0.600 &  0.038 $\pm$  0.015 $\pm$  0.015 &              &                                  \\
 0.600 - 0.800 &  0.014 $\pm$  0.005 $\pm$  0.006 &              &                                  \\
 0.800 - 1.000 & 0.0040 $\pm$ 0.0018 $\pm$ 0.0024 &              &                                  \\
\hline
\hline
\end{tabular}
\end{center}
\caption{\label{tab-fragaz} a$_0^{\pm}$ fragmentation function.
The quoted errors are statistical and systematic, respectively.
Because of the width of the \azpm, the relation between \xE\    and
\xp\   varies with mass.
Therefore the analysis is performed first with bins of \xE\  
(first two columns) and then repeated with bins of \lnxp.}
\end{table}

%========================================================================

\begin{table}[htb]
\begin{center}
\begin{tabular}{|c|c|c|c|c|c|}
\hline
 Particle &  \multicolumn{5}{c|}{  Location of the maximum
                                 of $d\sigma/d{\mathrm ln}(1/x_p)$ } \\
  \cline{2-6}
          & OPAL & L3~\protect\cite{bib-lpiz,bib-leta,bib-letap}
                 & DELPHI~\protect\cite{bib-delpiz}
                 & JETSET 7.4 & HERWIG 5.9 \\
\hline
 $\gamma$ & 4.61 $\pm$ 0.12 &           &      &     4.54   &    4.63    \\
  \piz\   & 3.77 $\pm$ 0.11 & 3.90 $\pm^{0.24}_{0.14}$ &
                              3.96 $\pm$ 0.13  &     3.78   &    3.86    \\
 $\eta$   & 2.64 $\pm$ 0.14 & 2.52 $\pm$ 0.12  &  &  2.94   &    3.01    \\
  \rpm\   & 2.63 $\pm$ 0.15 &                  &  &  2.69   &    2.70    \\
 $\omega$ & 2.89 $\pm$ 0.24 & 2.86 $\pm$ 0.20$^*$ &
                                               &     2.77   &    2.77    \\
  $\eta'$ & 2.21 $\pm$ 0.42 & 2.69 $\pm$ 0.10$^*$ &
                                               &     2.48   &    2.17    \\
  \azpm\  & 2.57 $\pm$ 0.50 &                  &  &  2.62   &    2.72    \\
\hline
\hline
\end{tabular}
\end{center}
\caption{\label{tab-lnxpmax}
Location of the maximum of the \lnxp\   distributions, determined from a
Gaussian fit to the data in the region around the maximum. 
The OPAL measurements are compared to other LEP measurements and to 
the predictions of the default versions of JETSET 7.4 and HERWIG 5.9.
The errors on the predictions are typically $\pm$0.01.
The results marked with an asterisk are
extracted assuming that the shape of the \lnxp\   distribution is
given by a MLLA calculation~\protect\cite{bib-letap}.}
\end{table}

%========================================================================

\begin{table}[htb]
\begin{center}
\begin{tabular}{|l|r|r|r|r|r|r|r|}
\hline
       &  $\gamma$ & \piz\   & $\eta$ & \rpm\   & $\omega$ & $\eta'$ & \azpm \\
\hline
\hline
  Measured range:        &      &      &      &      &      &      &      \\
  Min. \lnxp\            & 0.00 & 0.69 & 0.00 & 0.00 & 0.51 & 0.22 & 0.00 \\
  Max. \lnxp\            & 5.81 & 5.06 & 3.82 & 5.00 & 4.01 & 3.09 & 3.50 \\
\hline
 \% of rate in range   &      &      &      &      &      &      &      \\
  JETSET 7.4           & 81.9 & 88.7 & 78.7 & 98.9 & 87.7 & 68.9 & 80.5$^*$ \\
  HERWIG 5.9           & 79.4 & 87.7 & 77.9 & 98.7 & 87.1 & 74.5 & 78.6 \\
  Gaussian fit         & 78.7 & 84.9 & 82.2 & 97.4 & 82.3 & 77.7 & 81.6 \\
  $\pi^{\pm}$ shape    &      & 88.4 &      &      &      &      &      \\
  K$^0_{\mathrm{S}}$ shape
                       &      &      & 82.2 &      &      &      &      \\
\hline                 
 Combined \%           & 80.3 & 86.8 & 81.1 & 98.2 & 85.0 & 73.3 & 80.1 \\
 Error                 &  1.6 &  1.9 &  3.1 &  0.8 &  2.7 &  4.4 &  1.5 \\
\hline
 Measured rate        & 16.84 & 8.29 & 0.789 & 2.36 & 0.883 & 0.103 & 0.214 \\
 Extrapolated rate    & 20.97 & 9.55 & 0.973 & 2.40 & 1.039 & 0.141 & 0.267 \\
 Extrapolation error  &  0.42 & 0.21 & 0.038 & 0.02 & 0.033 & 0.008 & 0.005 \\
\hline
\hline
\end{tabular}
\end{center}
\caption{\label{tab-extrap}
Data used for the extrapolation to the unobserved energy/momentum
ranges. The result marked with an asterisk is from the JETSET tune of
ref.~\protect\cite{bib-tunenew}.}
\end{table}

%========================================================================

\begin{table}[htb]
\begin{center}
\begin{tabular}{|c|c|c|c|c|c|c|}
\hline
               & \multicolumn{4}{c|}{Experimental results} & JETSET & HERWIG \\
  \cline{2-5}
               & OPAL & ALEPH~\protect\cite{bib-alcomp}
                      & DELPHI~\protect\cite{bib-delpiz}
                      & L3~\protect\cite{bib-lpiz,bib-leta,bib-letap}
                      & 7.4 & 5.9 \\
\hline
\hline
     photon           &  & & & & & \\
   $x_E$ range        & 0.003-1.000
                      & 0.018-0.450 
                      & 
                      & & & \\
$N_{\gamma}$ in range &  16.84 $\pm$ 0.86
                      &   7.37 $\pm$ 0.24
                      &
                      & & & \\
$N_{\gamma}$ all $x_E$&  \gamrs
                      & 
                      &
                      & & 20.76 & 22.65 \\
\hline
\hline
     \piz\              &  & & & & & \\
   $x_E$ range        &  0.007-0.400
                      &  0.025-1.000
                      &  0.011-0.750
                      &  0.004-0.150 & & \\
 $N_{\pi^0}$ in range &  8.29 $\pm$ 0.63
                      &  4.80 $\pm$ 0.32
                      &  7.1  $\pm$ 0.8
                      &  8.38 $\pm$ 0.67 & & \\
$N_{\pi^0}$ all $x_E$ &  \pizrs
                      &  9.63 $\pm$ 0.64
                      &  9.2  $\pm$ 1.0
                      &  9.18 $\pm$ 0.73 & 9.60 & 10.29 \\
\hline
\hline
     $\eta$           &  & & & & & \\
    $x_E$ range       &  0.025-1.000
                      &  0.100-1.000
                      &
                      &  0.020-0.300 & & \\
 $N_{\eta}$ in range  &  0.79 $\pm$ 0.08
                      & 0.282 $\pm$ 0.022
                      &
                      &  0.70 $\pm$ 0.08 & & \\
$N_{\eta}$ all $x_E$  &  \etars 
                      &
                      &
                      &  0.91 $\pm$ 0.11 & 1.00  & 0.92  \\
 $N_{\eta}$ $x_p>0.1$ &  \etaresh
                      &  0.282 $\pm$ 0.022
                      &
                      &                 & 0.286 & 0.243 \\
\hline
\hline
 \rpm\                &  & & & & & \\
    $x_E$ range       &  0.016-1.000
                      &
                      &
                      & & & \\
$N_{\rho^{\pm}}$ in range &  2.36 $\pm$ 0.42
                      &
                      &
                      & & & \\
$N_{\rho^{\pm}}$ all $x_E$&  \rhors
                      &
                      &
                      & & 2.82 & 2.29 \\
\hline
\hline
 $\omega$             &  & & & & & \\
    $x_E$ range       &  0.025-0.800
                      &  0.053-1.000
                      &
                      &  0.026-0.300 & & \\
$N_{\omega}$ in range &  0.88 $\pm$ 0.12
                      &  0.64 $\pm$ 0.08
                      &
                      &  0.94 $\pm$ 0.14 & & \\
$N_{\omega}$ all $x_E$&  \omers
                      &  1.07 $\pm$ 0.14
                      &
                      &  1.17 $\pm$ 0.17 & 1.35 & 1.14 \\
\hline
\hline
   $\eta'$            &  & & & & & \\
    $x_E$ range       &  0.050-0.800
                      &  0.100-1.000
                      &
                      &  0.023-0.240 & & \\
$N_{\eta'}$ in range  & 0.103 $\pm$ 0.017
                      & 0.064 $\pm$ 0.014
                      &
                      &  & & \\
$N_{\eta'}$ all $x_E$ &  \etprs
                      &
                      &
                      &  0.25 $\pm$ 0.04 & 0.297 & 0.122 \\
$N_{\eta}'$ $x_p>0.1$ &  \etpresh
                      &  0.064 $\pm$ 0.014
                      &
                      &                & 0.127  & 0.060 \\
\hline
\hline
 \azpm\               &  & & & & & \\
    $x_E$ range       &  0.050-1.000
                      &
                      &
                      & & & \\
$N_{{\mathrm a}_{0}^{\pm}}$ in range &  0.21 $\pm$ 0.08
                      &
                      &
                      & & & \\
$N_{{\mathrm a}_{0}^{\pm}}$ all $x_E$&  \azers
                      &
                      &
                      & & 0.210 & 0.221 \\
\hline
\hline
\end{tabular}
\end{center}
\caption{\label{tab-others}
Summary of the measurements of the particles rates described
in this paper compared to other measurements at LEP and to JETSET
and HERWIG predictions.}
\end{table}

\clearpage
 
%========================================================================

%============================================ Figure 1
 
%============================================ Figure   2

%============================================ Figure 3

%============================================ Figure 4
 
%============================================ Figure  5

%============================================ Figure  6

%============================================ Figure  7

%============================================ Figure  8

%============================================ Figure  9

%============================================ Figure 10

%============================================ Figure 11

%============================================ Figure 12

%============================================ Figure 13

%============================================ Figure 14

%============================================ Figure 15

%========================================================================

\end{document}